\documentclass[5p,number,sort,compress,10pt,a4paper]{elsarticle}
\usepackage{amssymb}
\usepackage{mathrsfs,amsmath}
\usepackage{bm}
\usepackage[toc,page]{appendix}
\usepackage{natbib}
\usepackage{graphicx}
\usepackage{color,soul}
\usepackage{float}
\usepackage{todonotes}
\usepackage{makecell}
\usepackage{placeins}
\usepackage{caption}
\usepackage{subcaption}
\usepackage[export]{adjustbox}
\usepackage{algorithm}
\usepackage{algpseudocode}

\newcommand\blfootnote[1]{%
  \begingroup
  \renewcommand\thefootnote{}\footnote{#1}%
  \addtocounter{footnote}{-1}%
  \endgroup
}
\newcommand{\norm}[1]{\left\lVert#1\right\rVert}

\journal{tbd}

\begin{document}

\begin{frontmatter}

\title{Microstructural Generative Inverse Models for Additive Manufacturing Digital Twins}

\author[MSD]{Pablo Luna Falcon\corref{cor1}}\ead{lunafalconpa@ornl.gov}
\author[MSD]{Jason Mayeur}\ead{mayeurjr@ornl.gov}
\author[EEID]{Amirkoushyar Ziabari}\ead{ziabariak@ornl.gov}
\author[MSD]{Vincent Paquit}\ead{paquitvc@ornl.gov@ornl.gov}
\author[MSD]{Patxi Fernandez-Zelaia}\ead{fernandezzep@ornl.gov}

\address[MSD]{Manufacturing Science Division, Oak Ridge National Laboratory, Oak Ridge, TN, United States}
\address[EEID]{Electrification \& Energy Infrastructure Division, Oak Ridge National Laboratory, Oak Ridge, TN, United States}

\cortext[cor1]{Corresponding author}

\begin{abstract}

Metal additive manufacturing (AM) offers substantial advantages to the fabrication of complex structural components, but its widespread adoption is slowed due to complexities in the process-structure-property relations which is the cause for uncertainty and drives risk. Accurately quantifying these complex relations solely via experimental workflows is time consuming and costly. Integrated computational materials and engineering (ICME) frameworks address this challenge by leveraging advanced computational tools, such as crystal plasticity finite element models (CPFEM), to predict critical mechanical properties directly from microstructural data. However, these frameworks require a large datasets that are often difficult to obtain, and their predictive accuracy depends on the availability of realistic AM microstructure representations. Generative deep learning models are a promising approach towards synthesizing AM microstructures which contain complex morphological and crystallographic features. Denoising diffusion probabilistic models (DDPMs), in particular, show a strong potential to capture the inherent complexities of AM structures. In this work, we demonstrate that DDPMs can reliably synthesize digital-twin orientation-imaging micrographs, and we support this claim with quantitative evidence of the fidelity of the generated samples. Additionally, we establish a classifier-free guidance framework for solving complex inverse problems such as property-to-structure and process-to-structure mappings, and we showcase a potential application on a qualification-relevant uncertainty quantification problem. Our work highlights the potential of AI-based generative modeling as a key component of computational frameworks for accelerating AM materials qualification and certification. 
\end{abstract}


\end{frontmatter}

\blfootnote{\textbf{Notice of Copyright}. This manuscript has been authored by UT-Battelle, LLC under Contract No. DE-AC05-00OR22725 with the U.S. Department of Energy. The United States Government retains and the publisher, by accepting the article for publication, acknowledges that the United States Government retains a non-exclusive, paid-up, irrevocable, world-wide license to publish or reproduce the published form of this manuscript, or allow others to do so, for United States Government purposes. The Department of Energy will provide public access to these results of federally sponsored research in accordance with the DOE Public Access Plan (http://energy.gov/downloads/doe-public-access-plan).}

\section{Introduction}
\label{section:intro}

Metal additive manufacturing technologies aim to revolutionize the fabrication of complex structural components essential for energy, aerospace, automotive, and biomedical sectors. Applications include rapid prototyping, customization or repair of individual components, near-net-shape manufacturing, and functionally graded material fabrication. Furthermore, fusion based powder-bed AM processes offer opportunities to form material systems previously not possible to cast into complex geometries \cite{murr2010characterization,ledford2023microstructure,kareem2026additive, fernandez2021crystallographic}. However, widespread implementation of AM-processed components is limited by differences in mechanical behavior when compared to conventional, well-studied processing methods such as casting and forging. Despite crack-free dense material processing being already possible with AM for certain alloy systems, the process-structure-property relationships of critical mechanical properties (fatigue, creep, fracture, etc.) remain relatively unexplored. The inherently complex process-dependent thermomechanical histories and multi-scale physics increase the number of laborious and costly trial-and-error experiments that often fail to capture key physical phenomena.

\begin{figure}[t]
  \centering

  \includegraphics[width=1.0\linewidth, trim=1.2in 0.1in 1.5in 0.1in, clip]{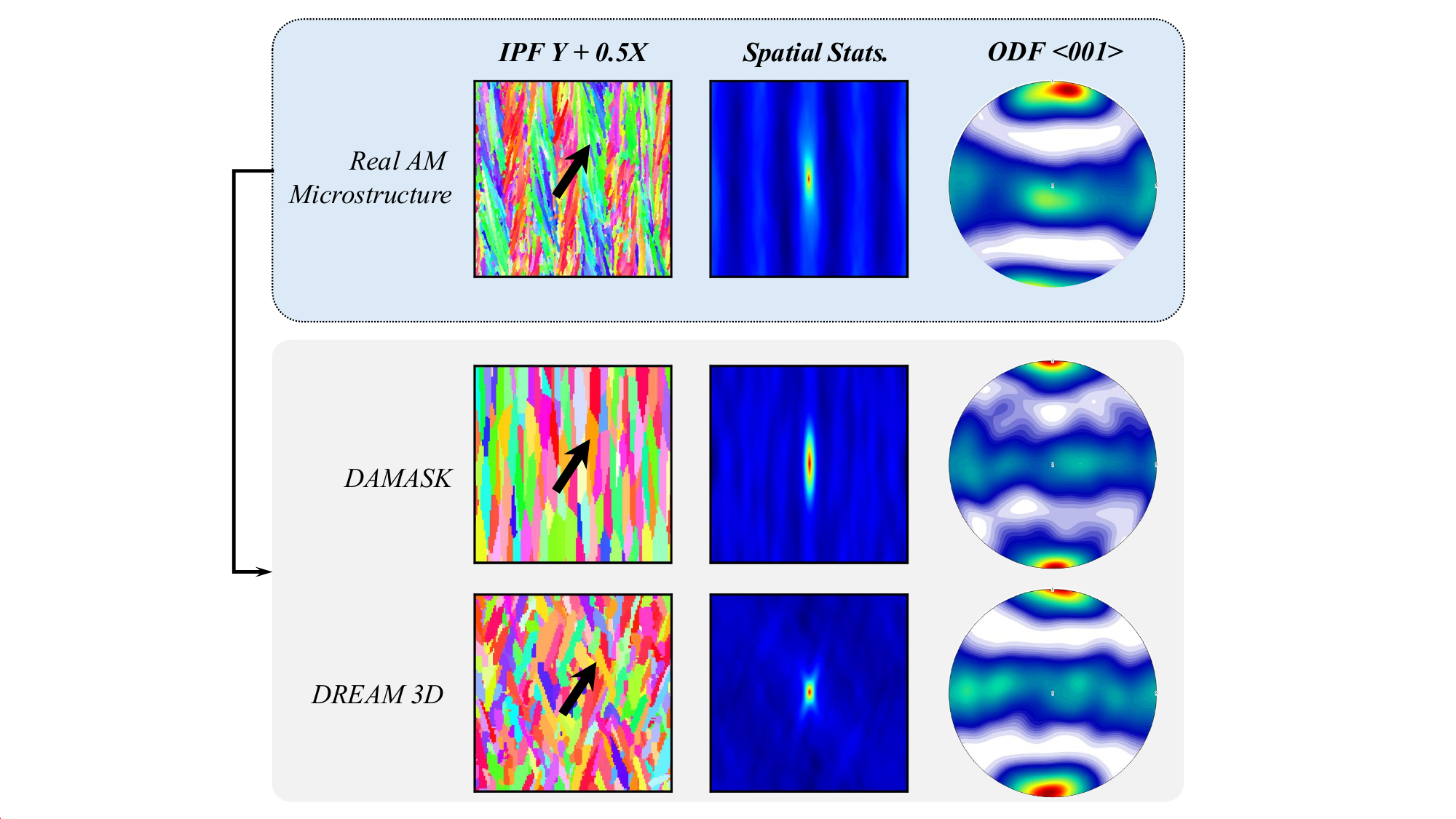}
  \caption{\textbf{Synthetic AM Microstructure Generation Strategies.} (Top) IN 718 microstructure fabricated via EB-PBF \cite{ raghavan2021influence}. (Bottom) Synthetic microstructures generated using DAMASK and DREAM.3D from summary statistics. The black arrow corresponds to the IPF plotting direction showing periodic texture bands which is representative of the complex texture characteristics of AM microstructures}
  \label{fig:motivation}
\end{figure}

Advances in accessibility to open source frameworks and an abundance of process data provide researchers with an assortment of flexible numerical tools to quantify AM process-structure-property relationships \cite{bozek2021nonlinear, scime2023data, villarraga2025multiscale}.  Grain-scale (meso-scale) models (phase-field, cellular automation, crystal plasticity, etc), are used to simulate microstructure evolution, fracture mechanisms, and mechanical behavior in metals \cite{steinbach2009phase}. Mesoscale features in structural metallic systems such as grain-size, crystallographic orientation, and spatial structure directly influence the aggregate material constitutive behavior and performance characteristics such as fatigue and creep \cite{mcdowell2009integrated,adams2012microstructure}.  While isotropic constitutive behavior is a valid assumption for equiaxed uniformly textured microstructures conventionally found in cast materials, AM materials are typically columnar with fiber textures. Highly textured materials have asymmetrical properties due to the inherent anisotropic nature of crystals \cite{roters2010overview}.  DED-fabricated Hastelloy C276 superalloy, for instance, exhibits a longer creep rupture time when loaded along the deposition direction than along the transverse direction \cite{qiu2024crystallographic}, and the tensile behavior of WA-DED SS316L and EBM-Inconel 718 with sharp texture are both highly dependent on the loading direction  \cite{thapliyal2024considering, kirka2017mechanical}.  Crystal Plasticity Finite Element Models (CPFEM) become critical when orientation-dependent properties govern the aggregate mechanical behavior of a material. These models quantify how mesoscale features influence the bulk mechanical properties, bridging polycrystalline microstructure information from the meso-scale up to the bulk scale \cite{kouraytem2021modeling, ozturk2018effect}. Critically, because the microstructure itself can be an input to the CPFEM model, realistic AM structure representations are required to accurately estimate the properties. 

Integrated Computational Materials and Engineering (ICME) approaches have been established to leverage numerical tools, such as CPFEM, to accelerate optimal parameter searching, material discovery, and component certification \cite{panchal2013key,furrer2023development}. Despite identical loading conditions, materials with microstructure-sensitive properties can exhibit vastly different behaviors. By leveraging grain-scale models, uncertainty can be quantified through simulations of statistically equivalent microstructures to extract descriptors capable of modeling the stochastic response \cite{mcdowell2007simulation, cheng2022effect, stopka2021effects}. This method requires an extensive collection of  Representative Value Elements (RVEs), which are often difficult to obtain. To address data scarcity, traditional ICME frameworks synthesize unique RVEs matching specific morphological or crystallographic descriptors using tools such as DREAM.3D, or Neper \cite{groeber2014dream, quey2011large}. However, these tools require the user to explicitly prescribe summary statistics to guide the RVE sampling process. Summary statistics often fail to capture complex geometric relationships and outliers which subsequently leads to a lack of realistic features within generated structures as seen in Fig.~\ref{fig:motivation}. There is yet to be a reliable generative tool for synthesizing realistic polycrystalline microstructures that effectively capture the inherent complexities of AM materials.

The materials community have leveraged generative deep learning models for inverse design frameworks, accelerating physics simulations, and synthetic data augmentation to name a few applications \cite{long2024generative, narikawa2022generative, zhang2018deep}. Three major architectures are commonly used for generation tasks, Variational Autoencoders (VAEs), Generative Adversarial Networks (GANs), and Denoising Diffusion Probabilistic Models (DDPMs) \cite{liu2023generative, yang2023two, kingma2013auto}. VAEs learn to compress high dimensional data into a probabilistic latent representation with an encoder and the network is then trained to reconstruct the input data from this latent space using a decoder \cite{kingma2013auto}. The compressed representation has been linked to mechanical properties enabling latent space exploration and optimization \cite{kim2021exploration}. VAEs however, are not ideal for image generation tasks due to the regularization term in the objective, which pushes the latent space to be normally distributed, and often leads to blurry and less detailed reconstructions \cite{bond2021deep}. GANs and DDPMs are generally preferred over VAEs for high-quality image generation tasks. GANs generate data by sampling from a latent space representation and mapping this to the appropriate data modality via the learned generator model \cite{goodfellow2014generative}. GANs have been proven useful to construct realistic samples for material simulations \cite{narikawa2022generative, kench2021generating, liu2020predicting, ziabari2025constrained, tabassum2024adapting}, as well as microstructure generation conditioned on processing parameters \cite{iyer2019conditional}. However, GANs are highly susceptible to mode collapse, where the generator stops capturing the diversity of the full training dataset and instead converges towards producing a subset type of image \cite{arjovsky2017towards}. Recently, DDPMs have surpassed GANs as state-of-the-art for image generation tasks \cite{dhariwal2021diffusion}. These models are flexible, scalable, expressive, and easy to train. They work by gradually corrupting the structure of the data distribution during the forward process, and then learn to restore structure over discrete steps in the reverse process \cite{ho2020denoising}. DDPMs bypass the limitations of GANs and VAEs by breaking down the generative process into a sequence of generative steps. A number of application cases have already been published for material science applications including molecular design \cite{wu2022diffusion}, alloy design \cite{fernandez4698278denoising}, and conditional micrograph synthesis \cite{azqadan2023predictive}. 

DDPMs have been applied for complex generative tasks such as anisotropic microstructure reconstruction, and multi-plane denoising for 2D-to-3D microstructure generation \cite{lee2024denoising, dureth2023conditional}. However, the majority of existing studies are primarily focused on generative tasks of binary phases, and in most cases, utilize datasets with relatively similar statistics which simplifies generation tasks  \cite{azqadan2023predictive, lyu2024microstructure, lee2024microstructure}. In polycrystalline microstructures, individual grains are characterized not only by morphology but also by a crystallographic orientation, which turns generation into a much more challenging task \cite{buzzy2026polymicros}.

This work demonstrates the use of DDPMs to generate statistically representative realizations of AM microstructures. The model is trained on a diverse dataset of complex polycrystalline orientation maps containing features commonly observed in AM materials. We introduce conditioning on a number of relevant variables to support property-to-structure and process-to-structure mapping. Analysis of the generated samples reveal that the model correctly captures the complex relationships among microstructural features, synthesizes highly realistic structures with greater fidelity than those generated by traditional methods, and consistently adheres to the specified conditioning information. Our aim is to feed these generated microstructures as inputs into downstream property simulators. Additionally, our results will assist in the acceleration of AM-processed component certification by advancing uncertainty quantification of critical mechanical properties via ICME frameworks, and by revealing structure-property relationships through conditional generation.

\section{Methods}
\label{section:methods}

\begin{figure}[t]
  \centering

  \includegraphics[width=1.0\linewidth, trim=1.2in 0.1in 1.5in 0.1in, clip]{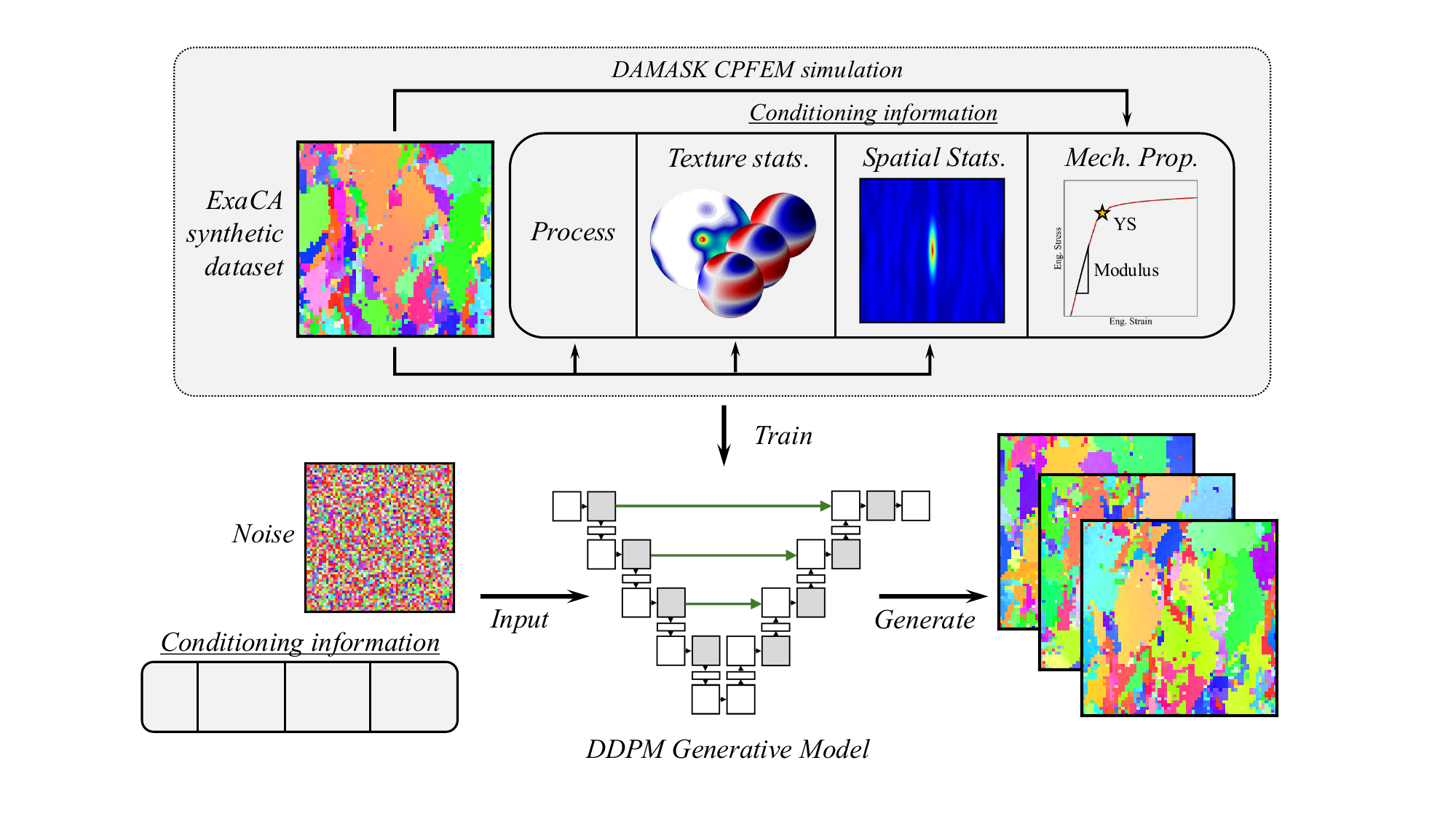}
  \caption{\textbf{Overall Workflow Schematic.} The DDPM was trained with $64^2$ arrays of \textit{exaCA} simulated EBSD micrographs. Generation may be conditioned on process, texture, morphology and mechanical properties.}
  \label{fig:workflow}
\end{figure}

The DDPM was trained on Electron Backscatter Diffraction (EBSD) micrographs where the crystal orientation $g$ is quantified at each spatial location using Bunge-Euler angles, $g = (\phi_1, \Phi, \phi_2)$. Training data was collected from a freely available AM synthetic polycrystalline dataset \textit{exaCA} \cite{knapp2025simulated}. This data was generated via Cellular Automation (CA) and consists of six unique RVEs, each corresponding to different AM processing parameters (power, speed, and hatch spacing), of size $401 \times 401 \times 577$. The RVEs were randomly cropped into 1500 sub-samples of $64^3$ arrays and randomly rotated with the crystallographic orientations updated accordingly. Training was done with randomly sampled $64^2$ XY-slices, and subsequently augmented with random rotations, horizontal flips, and vertical flips to increase data diversity. Each example was paired with summary statistics describing morphology, crystallographic texture, process, and mechanical properties (yield stress and uniaxial elastic modulus). Morphological information is captured using orientation imaging two-point correlation functions \cite{fernandez2019process, fernandez2021nickel}. Texture, which is mathematically described as an Orientation Distribution Function (ODF), is represented as a Generalized Spherical Harmonics (GSH) series expansion truncated up to $l=4$ and split into the real and imaginary components \cite{bunge2013texture, buzzy2024statistically}. Process is represented using a one-hot encoding corresponding to the six \textit{exaCA} RVE classes. Lastly, mechanical properties, in this case the uniaxial elastic modulus and initial yield strength, were estimated from simulations performed using open source code DAMASK \cite{roters2019damask}. During evaluation, the DDPM-generated samples can be biased to exhibit specific qualities (properties, texture, morphology, etc.) via introduction of these four quantities as conditioning information. Details will follow in the DDPM section. The overall training and generation workflow is shown in Fig. \ref{fig:workflow}.

\subsection{Quantitative Texture Descriptor}
\label{ssection:gsh}
An EBSD micrograph consists of spatially indexed crystal orientations quantified using Bunge-Euler notation $\mathbf{g}$ defined over a Fundamental Zone $(FZ)$. Crystallographic texture is a statistical description, modeled by an ODF $f(\mathbf{g})$, that maps how likely a crystal is oriented in any direction relative to a reference coordinate system. Similar to any 2D periodic function, which can be approximated by a weighed sum of sines and cosines, 3D periodic functions, such as ODFs, can be approximated by a weighed sum of Generalized Spherical Harmonics (GSH). The formulation described by Bunge \cite{bunge2013texture} encodes the redundancies of crystal symmetries as follows,

\begin{equation}
\begin{aligned}
f(\mathbf{g}) = \sum^\infty_{l=0}\sum^{M(l)}_{\mu=1}\sum^{+l}_{n=-l}F^{\mu n}_l\overset{:\,}{T}{^{\mu n}_l}(\mathbf{g})
\label{eqn:harmonic_expansion},
\end{aligned}
\end{equation}

\noindent where $F^{\mu n}_l$ and $\overset{:\,}{T}{^{\mu n}_l}$ are the complex valued expansion coefficients and GSH basis functions symmetrized for cubic crystal symmetry and no sample symmetry. Analogously to the way how Fourier coefficients can be determined, the coefficients $F^{\mu n}_l$ can be obtained by exploiting the orthogonality of $\overset{:\,}{T}{^{\mu n}_l}$,

\begin{equation}
\begin{aligned}
F^{\mu n}_{l} = (2l+1)\int f(\mathbf{g}) \overset{:\,}{T}{^{\mu n}_l}(\mathbf{g}) d\mathbf{g}
\label{eqn:gsh_coefficient},
\end{aligned}
\end{equation}

A vector with the real and imaginary components of $F^{\mu n}_l$ can then be used as a quantitative metric to describe texture \cite{fullwood2010microstructure}.

\subsection{Morphological Statistics}
\label{ssection:spatial_stats}
Two-point correlation functions $p(\mathbf{g}, \mathbf{g}^\prime | \mathbf{t})$ are used to describe the overall grain morphology of a microstructure by calculating the spatial correlations of crystallographic orientations between two points across all coordinates, where $\mathbf{g}$ and $\mathbf{g}^\prime$ correspond to the material state at the tail and head of a vector $\mathbf{t}$. Discrete material states, such as binary phase maps, can be simply encoded as phase 0 and phase 1.  However, crystallographic orientation is a continuous variable which complicates the description of the material state. The following formulation is required,

\begin{equation}
\begin{aligned}
p(\mathbf{g},\mathbf{g}^\prime|\mathbf{t}) = \frac{1}{|\mathcal{X}|}\int_\mathcal{X}f_x(\mathbf{g})f_{x+t}(\mathbf{g^\prime})d\mathcal{X}
\label{eqn:two-point},
\end{aligned}
\end{equation}

\noindent where $\mathcal{X}$ is the total number of pixels. If $\mathbf{g} = \mathbf{g}^\prime$ then this quantity describes the \textit{autocorrelation}. Averaging the spatial autocorrelation over all possible $\mathbf{g}$, while losing information about absolute texture such as fiber type and direction, retains structural morphological information,

\begin{equation}
\begin{aligned}
\bar{p}_t = \frac{1}{V_{FZ}}\int_{FZ}p(\mathbf{g},\mathbf{g}^\prime|\mathbf{t})d\mathbf{g}
\label{eqn:integrate_ori},
\end{aligned}
\end{equation}

\noindent where $V_{FZ}$ is the fundamental zone volume. The fundamental zone is the reduced region in orientation space $\mathbf{g}$ that contain all physically distinct orientations under the crystal symmetry. By using a the GSH representation of $f_x(\mathbf{g})$ described in Eq. \ref{eqn:harmonic_expansion}, and by combining Eq. \ref{eqn:two-point}  and Eq. \ref{eqn:integrate_ori}, the following expression can be derived,
\begin{equation}
\begin{aligned}
\bar{p}_t 
&= \frac{1}{|\mathcal{X}|} \frac{1}{V_{FZ}} \int\limits_{\mathcal{X}}\int\limits_{V_{FZ}} f_x(\mathbf{g})f^{*}_{x+t}(\mathbf{g})d\mathbf{g}d\mathcal{X}\\
&= \frac{1}{|\mathcal{X}|} \frac{1}{V_{FZ}} \int\limits_{\mathcal{X}}\int\limits_{V_{FZ}} \sum_{\mu, n,l}F^{\mu n}_{l x}\overset{:\,}{T}{^{\mu n}_l}(\mathbf{g}) \sum_{\mu, n, l}F^{\mu n *}_{l x+t}\overset{:\,}{T}{^{\mu n *}_l}(\mathbf{g}) d\mathbf{g}d\mathcal{X}\\
&= \frac{1}{|\mathcal{X}|} \frac{1}{V_{FZ}} \frac{1}{2l+1}\int\limits_{\mathcal{X}} \sum_{\mu, n,l}F^{\mu n}_{l x}F^{\mu n*}_{l x+t} d\mathcal{X}\\
&= \frac{1}{|\mathbf{S}|} \frac{1}{V_{FZ}} \frac{1}{2l+1}\sum^\mathbf{S}_{s=1}\sum_{\mu, n, l} F^{\mu n}_{l s}F^{\mu n*}_{l s+t}
\label{eqn:spatial_stats_eq},
\end{aligned}
\end{equation}
\noindent where $f^*_{x+t}(\mathbf{g})$ is the complex conjugate. The expression can be significantly simplified by taking advantage of the orthogonality of GSH functions, which eliminates all terms except those with matching $l$, $\mu$, $n$. Lastly Eq. \ref{eqn:spatial_stats_eq} is computed as follows,
\begin{equation}
\begin{aligned}
\bar{p}_t = \frac{1}{|\mathbf{S}|}\frac{1}{V_{FZ}}\frac{1}{2l+1}\sum_{\mu,n,l}\mathscr{F}^{-1}(|\mathscr{{F}}(F^{\mu n}_l|^2)
\label{eqn:p_t},
\end{aligned}
\end{equation}

\noindent where $\mathscr{{F}}$ is the two-dimensional FFT and $\mathscr{{F}}^{-1}$ is the corresponding inverse transform. In short, the procedure consists of transforming the Euler angles into GSH coefficient maps, convolve each GSH coefficient with each other for all of the basis weights and lastly average them \cite{paulson2017reduced}. Autocorrelation maps capture structural patterns in microstrucures, in Fig.  \ref{fig:motivation} for instance, the alternating stripes of $<001>$ and $<011>$ causes the repeated bands in the map (spatial-orientation structure). For a detailed description regarding autocorrelation map interpretation refer to Fig. \ref{fig:ss_example}.

\begin{figure}[t]
  \centering
  \includegraphics[width=1.0\linewidth]{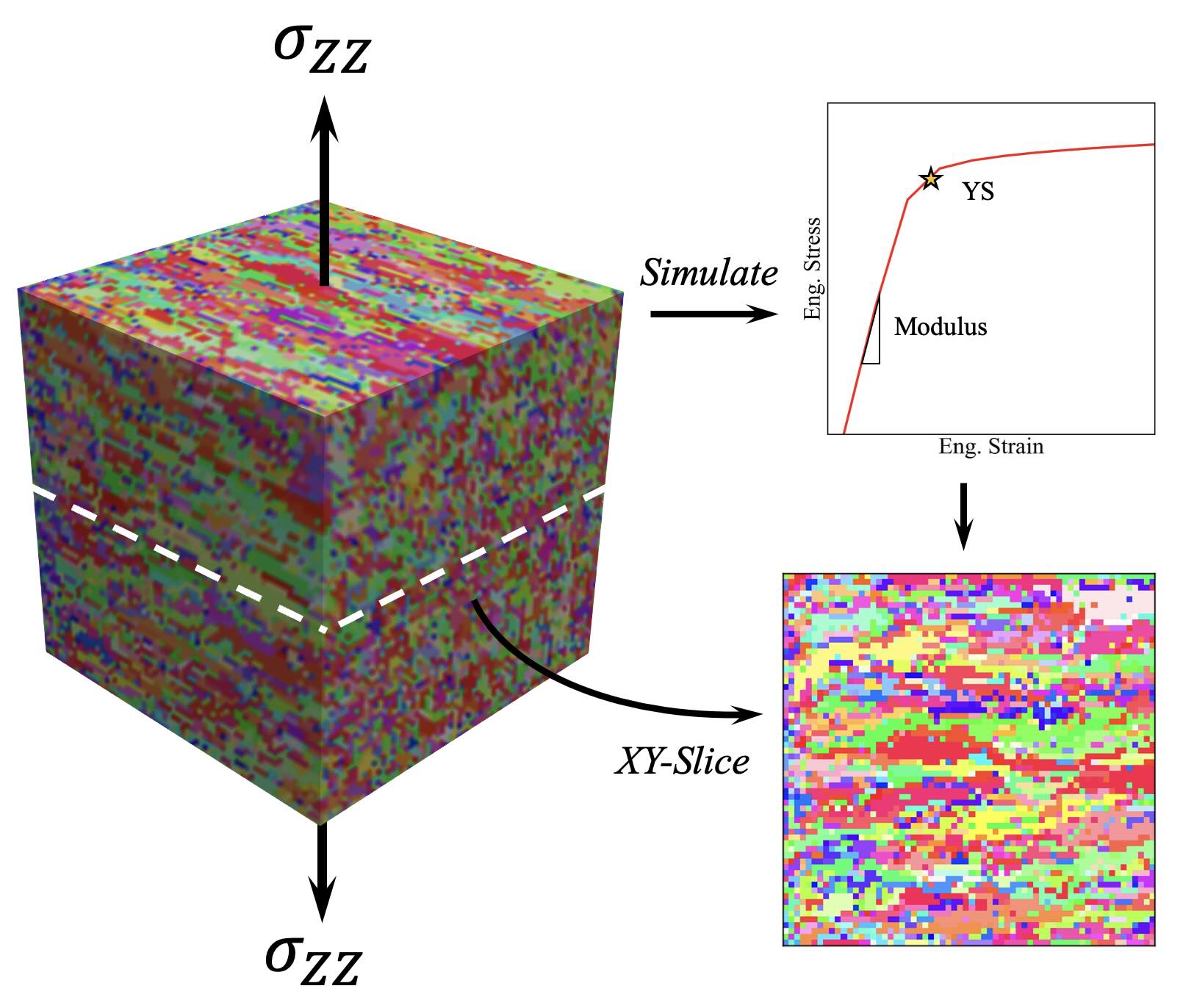}
  \caption{\textbf{Training Data Example.} Dataset consisted of $64 \times 64 \times 64$ array of crystallographic orientations quantified as Euler angles. Uniaxial tensile CPFEM simulations was performed on 3D blocks along the Z-axis. The respective bulk mechanical properties was used as property descriptors for every extracted XY-slice.}
  \label{fig:data_example}
\end{figure}

\subsection{Crystal Plasticity Simulations}
\label{ssection:CPFEM}

The grain scale deformation of an RVE under specific loading conditions can be simulated with a crystal plasticity model given knowledge of the representative grain structure and the crystallographic texture. DAMASK uses a finite deformation kinematic framework based on a multiplicative decomposition of the deformation gradient, $\mathbf{F}$, into elastic and plastic components,

\begin{equation}
\begin{aligned}
\mathbf{F}=\mathbf{F}^e\mathbf{F}^p
\label{eqn:def_gradient},
\end{aligned}
\end{equation}

\noindent where $\mathbf{F}_e$ and $\mathbf{F}_p$ are the elastic and plastic components, respectively. The plastic deformation evolves according to the velocity gradient in the isoclinic lattice configuration $\bar{\mathbf{L}}_p$,

\begin{equation}
\begin{aligned}
\bar{\mathbf{L}}^p = \dot{\mathbf{F}}^p\mathbf{F}^{p-1}
\label{eqn:velocity_gradient},
\end{aligned}
\end{equation}
\noindent The plastic velocity gradient is defined as the sum of slip rates over all active slip systems, denoted by $\alpha$, and is defined as follows,

\begin{equation}
\begin{aligned}
\bar{\mathbf{L}}^p = \sum_\alpha\dot{\gamma}^\alpha\left(\mathbf{b}^\alpha\otimes\mathbf{n}^\alpha \right) 
\label{eqn:plastic_velocity_gradient},
\end{aligned}
\end{equation}
\noindent where $\dot{\gamma}^\alpha$ is the slip rate, $\mathbf{b}^\alpha$ and $\mathbf{n}^\alpha$ are the unit vectors along the slip direction and the slip plane normal, respectively, in the reference configuration. A phenomenological power law available is used to model the slip kinetics,

\begin{equation}
\begin{aligned}
\dot{\gamma}^\alpha = \dot{\gamma}_0\left|{\frac{\tau^\alpha}{s^\alpha}}\right|^n \mathrm{sgn}(\tau^\alpha)
\label{eqn:shear_rate},
\end{aligned}
\end{equation}
where $s^\alpha$ is the slip resistance, $\tau^\alpha$ is the resolved shear stress, $\dot{\gamma}_0$ is the reference slip rate, and $n$ is the rate-sensitivity exponent. The slip resistance evolution is modeled with the following expression,

\begin{equation}
\begin{aligned}
\dot{s}^\alpha = \sum_\beta{h^{\alpha\beta}\left| \dot{\gamma}^\beta \right|}\\
h^{\alpha\beta}=q^{\alpha\beta}h_0\left(1-\frac{s^\beta}{s^s} \right)^m
\label{eqn:hardening_model},
\end{aligned}
\end{equation}

where $h^{\alpha\beta}$ is the hardening matrix that determines how slip on system $\beta$ influences slip resistance on system $\alpha$. The hardening matrix depends on a slip system interaction matrix $q^{\alpha\beta}$, the initial hardening modulus $h_0$, the saturation stress $s_s$, and the hardening exponent $m$. Simulations for all $64^3$ elements with uniaxial loading conditions applied along the z-axis, at a strain rate of $5\times10^{-3} s^{-1}$ over 100 time steps until reaching an applied strain of $5$ \%.  The corresponding uniaxial elastic modulus and yield strength extracted from the bulk response were paired with each XY-slice as shown in Fig. \ref{fig:data_example}.

\subsection{Denoising Diffusion Probabilistic Models}
\label{section:micro_stats}

The objective of a generative probabilistic model is to approximate the true underlying data probability distribution $q(x_0)$ by maximizing the likelihood of the training dataset. DDPMs model probability distributions by learning paths to reverse a diffusion process $p_\theta(x_{0:T})$ across a sequence of steps $t \in \{0,1,2,...,T \}$. A \textit{denoiser} neural network parametrizes a Markov process which gradually transforms gaussian noise $\mathbf{x}_T\sim\mathcal{N}\left(0,\mathbf{I}\right)$ into a likely sample $\mathbf{x}_0$ with simple transitions modeled as conditional gaussian distributions $p_\theta(\mathbf{x}_{t-1}|\mathbf{x}_t)$.  From a score-matching perspective, the model learns a time-dependent vector field $\mathbf{s}_t(\mathbf{x})$, known as the score, that points in directions in data space that move samples towards high-probability regions. Reversing diffusion can be understood as a random sample traversing data space through small stochastic steps, gradually moving toward regions of higher probability under a learned data distribution,

\begin{equation}
\begin{aligned}
p_\theta(\mathbf{x}_{t-1}|\mathbf{x}_t) := \mathcal{N}\left(\mathbf{x}_{t-1}; \bm{\mu}_\theta(\mathbf{x}_t,t), \beta_t\mathbf{I} \right)
\label{eqn:reverse_process},
\end{aligned}
\end{equation}

\noindent where $\mu_\theta$ is learned with a neural network, and $\beta_t$ is a fixed variance schedule following the original DDPM formulation \cite{ho2020denoising} . The diffusion forward process  $q(x_{1:T} | x_{0})$, is defined as a fixed Markov process consisting of the gradual addition of gaussian noise, ultimately approaching a gaussian distribution. The forward gaussian transitions are modeled as follows,

\begin{equation}
\begin{aligned}
q(\mathbf{x}_t|\mathbf{x}_{t-1}) := \mathcal{N}(\mathbf{x}_t;\sqrt{1-\beta_t}\mathbf{x}_{t-1}, \beta_t\mathbf{I)}
\label{eqn:forward_process},
\end{aligned}
\end{equation}

\noindent from which any arbitrary time step can be easily sampled as $\mathbf{x}_t =\sqrt{\bar{\alpha}}\,\mathbf{x}_0 + \sqrt{1-\bar{\alpha}}\,\bm{\epsilon}$ where $\bm{\epsilon} \sim \mathcal{N}\left(0,\mathbf{I}\right)$ and $\bar{\alpha}_t = \prod_{s=1}^t (1 - \beta_s)$. Notably, the reverse diffusion processes is tractable and has a closed form expression but only when conditioned on the original image $\mathbf{x}_0$,

\begin{equation}
\begin{aligned}
q(\mathbf{x}_{t-1}|\mathbf{x}_t,\mathbf{x}_0) := \mathcal{N}\left(\mathbf{x}_{t-1}; \bm{\tilde{\mu}}_\theta(\mathbf{x}_t,\mathbf{x}_0), \beta_t\mathbf{I} \right)
\label{eqn:reverse_process_conditioned},
\end{aligned}
\end{equation}

\begin{figure}[t]
  \centering
  \includegraphics[width=1.0\linewidth, trim=0.1in 0.1in 0.1in 0.1in, clip]{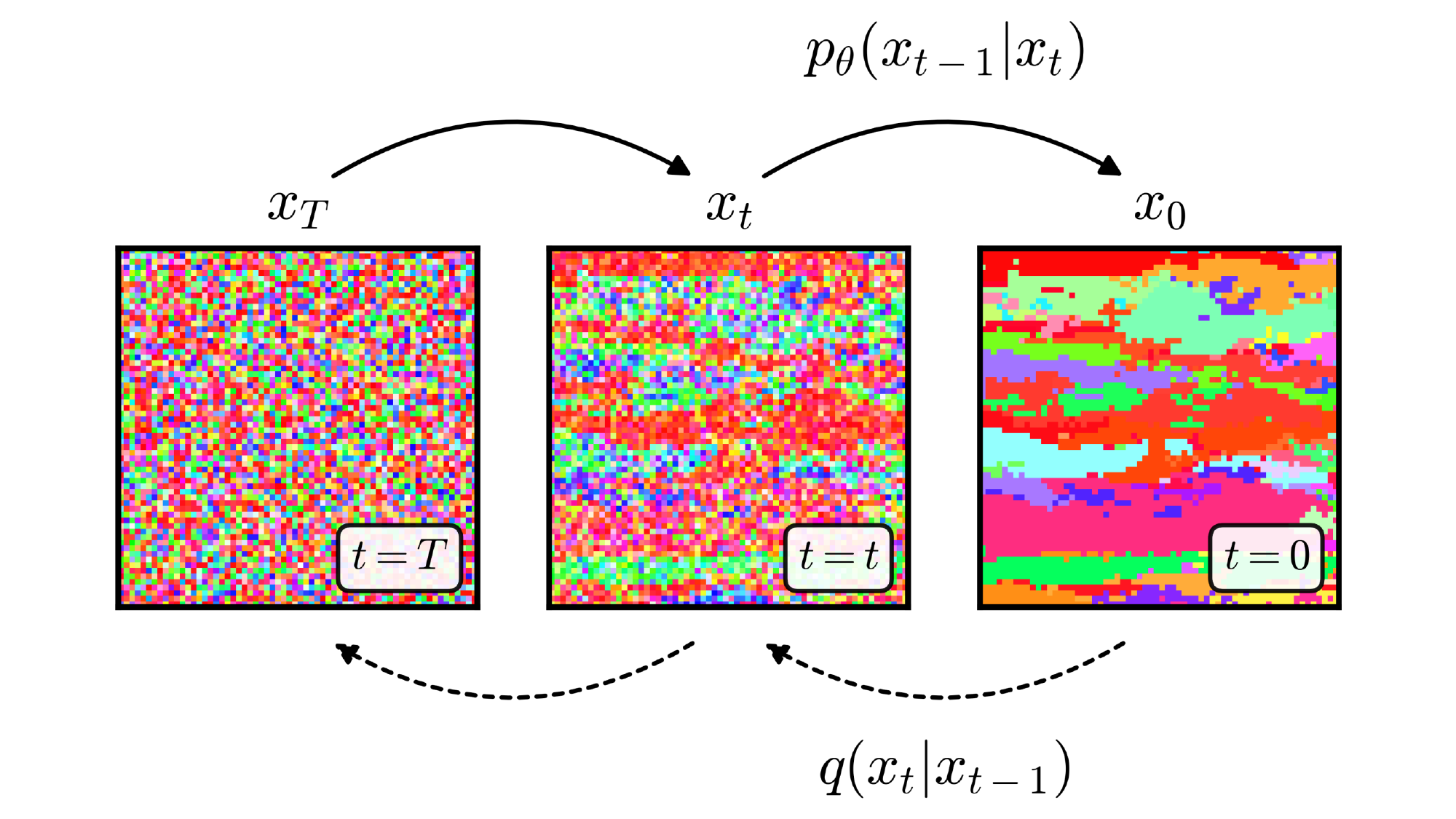}
  \caption{Forward (from right to left) and reverse (from left to right) diffusion process.}
  \label{fig:ddpm-for-rev}
\end{figure}

\noindent Both the forward and the reverse processes are depicted in Fig. \ref{fig:ddpm-for-rev}. A neural network trained on predicting the injected gaussian noise $\bm{\epsilon}_\theta(\mathbf{x}_t,t)$  from the forward process can sequentially denoise a sample. Training is done via variational inference by learning the reverse transition distributions $p_\theta(\mathbf{x}_{t-1}|\mathbf{x}_t)$, which approximates the true reverse transitions $q(\mathbf{x}_{t-1}|\mathbf{x}_t,\mathbf{x}_0)$ across all observed data and at all time steps $t$. The training objective can be simplified to a noise-prediction loss  $\norm{\bm{\epsilon}-\bm{\epsilon}_{\theta}\left(\bm{x}_t,t\right)}^2$ for arbitrary $t$ . A noisy sample $\mathbf{x}_{t}$ can be denoised to a corresponding previous step $\mathbf{x}_{t-1}$ using the following expression,

\begin{equation}
\begin{aligned}
\mathbf{x}_{t-1} = \frac{1}{\sqrt{\bar{\alpha_t}}}\left(\mathbf{x}_t - \frac{\beta_t}{\sqrt{1-\bar\alpha_t}}\bm{\epsilon}_\theta\left(\mathbf{x}_t,t \right)\right) + \sqrt{\beta_t}\,\mathbf{z}
\label{eqn:denoising_step},
\end{aligned}
\end{equation}

\noindent where $\mathbf{z} \sim \mathcal{N} \left(0,\mathbf{I}\right)$ and $\bar\alpha_t:=\prod^t_{s=1}(1-\beta_s)$. At the last denoising step $p\left(\mathbf{x}_0|\mathbf{x}_1\right)$, $\mathbf{z}=0$ since $\mathbf{x}_0$ represents a noise-free sample. To summarize, a DDPM generates data by learning stochastic trajectories in data space that map randomly sampled gaussian noise to realistic samples resembling those in the training dataset.

\begin{figure*}[t]
  \centering
  \includegraphics[width=1.0\linewidth, trim=0.5in 1.3in 0.5in 1.3in, clip]{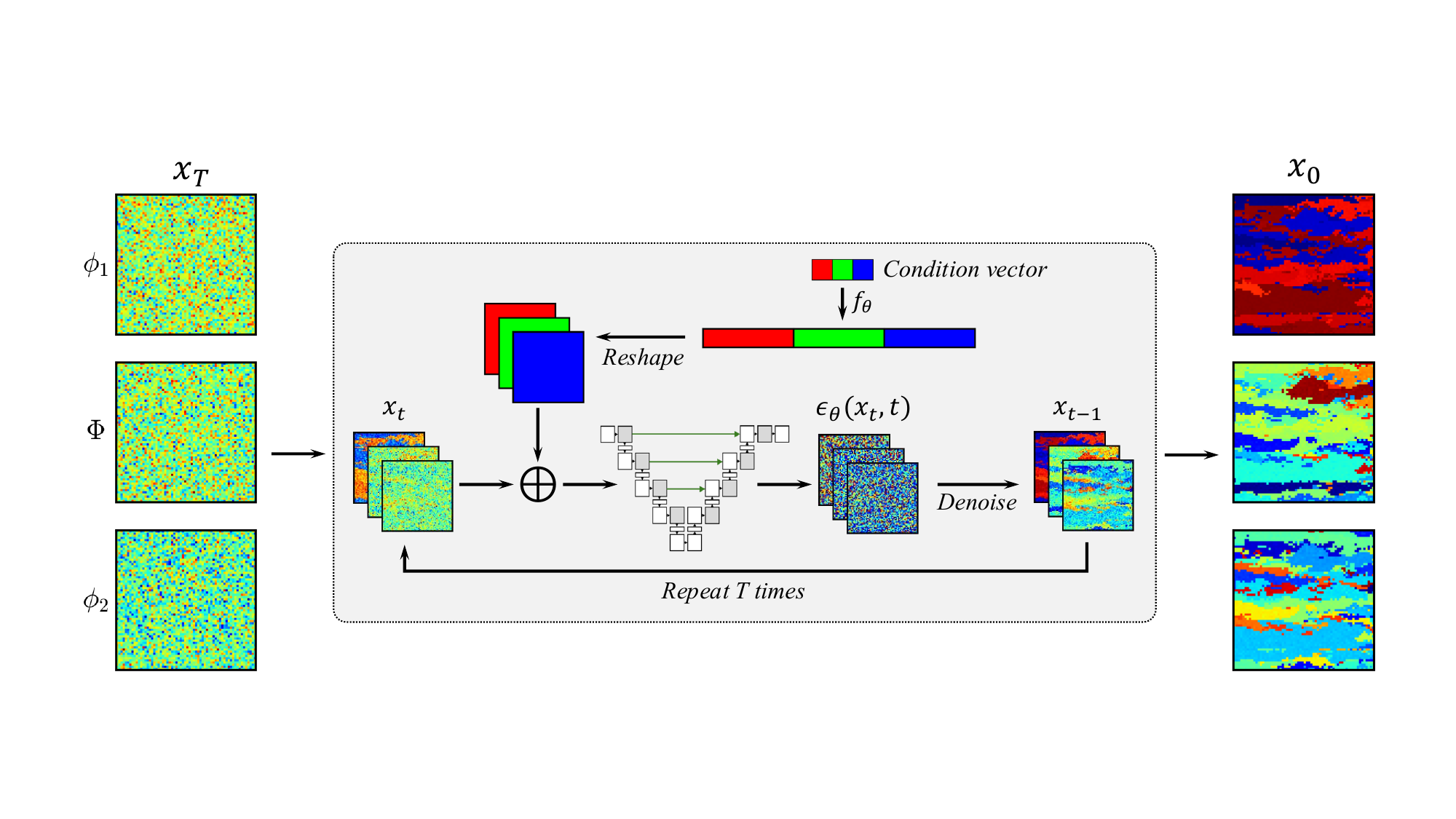}
  \caption{\textbf{Denoiser Architecture.} The network is a U-Net that takes in $x_t$ as input, which consists of a $64 \times 64 \times3$ Euler angle array with its respective conditioning information injected. Denoising is performed over 1000 time steps.}
  \label{fig:denoiser_architecture}
\end{figure*}

\subsection{Conditional Generation}
\label{section:conditional_gen}

A DDPM is free to stochastically sample data at any learned high probability regions. However, in some cases we may want to bias sampling to match certain characteristics, such as a particular texture or morphology. During training, contextual information $c$ is randomly provided with $p_{cond}$, encouraging the model to learn both conditional and unconditional distribution. During generation, an unconditionally sampled diffusion trajectory is guided by $\mathbf{c}$ to move toward regions in data space that match the desired characteristics. This strategy is called Classifier-Free Guidance (CFG) \cite{ho2022classifier} and is implemented during inference as follows,

\begin{equation}
\begin{aligned}
\bm{\tilde{\epsilon}}_\theta\left(\mathbf{x}_t,t,\mathbf{c}\right) = (1-w)\bm{\epsilon}_\theta\left(\mathbf{x}_t,t, \varnothing \right) + w\bm{\epsilon}_\theta\left(\mathbf{x}_t,t,\mathbf{c}\right) 
\label{eqn:conditional_eqn},
\end{aligned}
\end{equation}

\noindent which consists of a linear combination of predicted noise under a conditional distribution $\bm{\epsilon}_\theta\left(\mathbf{x}_t,t,\mathbf{c}\right)$ and the predicted noise under an unconditional distribution $\bm{\epsilon}_\theta\left(\mathbf{x}_t,t, \varnothing \right)$ at each time $t$. The guidance strength is controlled by $w$, where $w=0$ corresponds to unconditional generation, and $w=1$ to conditional generation, Notably, $w$ is unbounded, while any value between $0$ and $1$ represent an interpolation between unconditional and conditional generation, $w>1$ results in an overemphasis on matching conditioning information. 

\begin{algorithm}
\caption{Training}\label{alg:training}
\begin{algorithmic}
\State repeat
\State sample $\mathbf{x}_0 \sim q(\mathbf{x}_0)$
\State sample $t \sim Unif(\{1,...,T\})$
\State sample $\bm{\epsilon} \sim \mathcal{N}(\bm{0},\mathbf{I})$
\State Compute loss $L=\norm{ \bm{\epsilon} - \bm{\tilde{\epsilon}}_\theta\left(\mathbf{x}_t, t, \mathbf{c} \right) }^2 $
\State Compute gradient descent step with $\nabla_{\theta} L$
\end{algorithmic}
\end{algorithm}

\begin{algorithm}
\caption{Sampling}\label{alg:gen}
\begin{algorithmic}
\State sample $\mathbf{x}_T \sim \mathcal{N}(\bm{0},\mathbf{I})$
\For{t=T,...,1}
    \For{n=1,....N}
        \State sample $\mathbf{z}\sim \mathcal{N}(0,\mathbf{I})$ if $t>1$ else $\mathbf{z}=0$
        \State $\mathbf{x}_{t-1} = \frac{1}{\sqrt{\alpha_t}} \left(\mathbf{x}_t - \frac{1-\alpha_t}{\sqrt{1-\bar{\alpha}}_t} \bm{\tilde{\epsilon}}_\theta\left(\mathbf{x}_t,t, \mathbf{c}\right)\right) + \sqrt{\beta_t} \mathbf{z}$
        
        \If{$u<U$ and $t>1$}
        \State $\mathbf{x}_t \sim \mathcal{N}\left(\sqrt{1-\beta_{t-1}}\mathbf{x}_{t-1}, \beta_{t-1}\bm{I} \right)$
        \EndIf
    \EndFor
\EndFor
\State return $\mathbf{x}_0$
\end{algorithmic}
\end{algorithm}

The training and sampling algorithms are shown in Algorithm 1 and Algorithm 2 respectively. If needed, resampling steps can be incorporated over $N$ time steps if reharmonization is needed. Resampling consists of sequential sample denoising and re-noising at a single time step $t$, allowing the model to correct small numerical errors introduced by each denoising step. Resampling has been shown to improve the semantic quality of filled-in regions generated in an image \cite{lugmayr2022repaint}.  In material science, resampling has been used to mitigate discrepancies between generated layers for a 2D-to-3D microstructure reconstruction \cite{lee2024multi}. 

\subsection{Denoiser Architecture}
\label{section:denoiser_arch}

The denoising architecture is outlined in Fig.~\ref{fig:denoiser_architecture}. Our denoiser consists of a U-net architecture with each layer composed of residual and attention blocks. The model takes as input Euler angle maps reduced to their respective fundamental zones to eliminate redundancies in orientation space. 

\begin{figure}[t]
  \centering
  \includegraphics[width=1.0\linewidth]{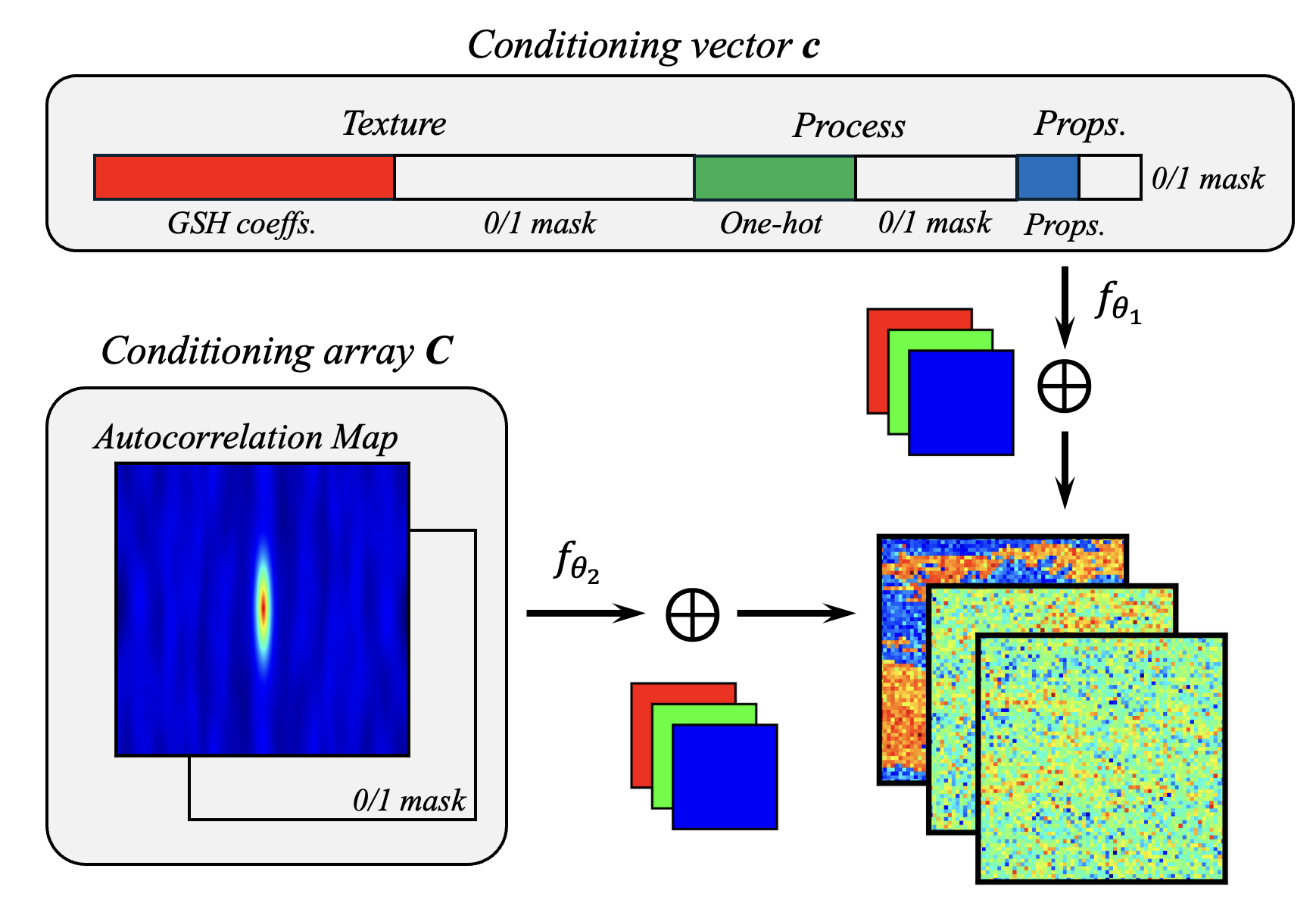}
  \caption{\textbf{Conditioning Strategy.} Conditioning information is broken down into a vector $\mathbf{c}$ containing texture as GSH coefficients, process as one-hot-encoding, and properties as yield and modulus. Each condition has a 0/1 vector concatenated that correspond do provided or omitted condition respectively. Since autocorrelation maps are $64 \times 64$ arrays, they are stored as a separate array $\mathbf{C}$ with its corresponding 0/1 array. Conditioning information is mapped to the same dimensionality as the data and added.}
  \label{fig:conditioning_vector}
\end{figure}

An overview of how the conditioning vector $\mathbf{c}$ and array $\mathbf{C}$ are structured is shown in Fig. \ref{fig:conditioning_vector}. Texture condition is a $20$-dimensional vector, corresponding to the real and imaginary components of the GSH coefficients. Process condition is a $6$-dimensional vector containing the one-hot encoding associated with each of the \textit{exaCA} RVEs generated under different process settings. Property condition is a $2$-dimensional vector containing the yield strength and the uniaxial elastic modulus. Lastly, morphology is a $64 \times 64$ array corresponding to the autocorrelation map. Individual conditions can be included or omitted by concatenating either a one-filled or zero-filled vector, respectively. For example, if conditional generation with texture is requested, the corresponding $20$-dimensional vector is concatenated with a $20$-dimensional vector of ones to form a $40$-dimensional vector, while the remaining conditions are filled with zeros. The full conditioning vector $\mathbf{c}$, which includes texture, process, and property information is formed by concatenating all three vectors with their respective mask. Morphology follows a similar strategy, with a one-filled or zero- filled $64 \times 64$ mask concatenated with the autocorrelation map to form $64 \times 64 \times 2$ array $\mathbf{C}$. While multiple methods of conditioning information injection exist, such as feature-wise linear modulation (FiLM), where auxiliary networks take in conditioning vectors as inputs to shift and scale activations \cite{perez2018film}, we opted for a simple additive conditioning,

\begin{equation}
\begin{aligned}
\mathbf{h^\prime} = \mathbf{h} + f_{\theta_1}(\mathbf{c}, t) + f_{\theta_2}(\mathbf{C},t)
\label{eqn:additive_cond},
\end{aligned}
\end{equation}
where $\mathbf{h}$ and $\mathbf{h}^\prime$ correspond to the activation and shifted activation values, respectively, $f_{\theta_1}$ is an MLP for vector-to-vector mapping, and $f_{\theta_2}$ is a CNN-to-MLP for array-to-vector mapping. At a given time $t$, they project $\mathbf{c}$ and $\mathbf{C}$ into the same dimensional space as $\mathbf{h}$. Conditioning information is injected at every U-Net layer. Training of a denoising process of $T=1000$ time steps was performed on eight 8GB A100 GPUs using a training dataset of $1500$ maps and an 80/20 training-validation split. Optimization was performed using Adam with a batch size of $32$,  learning rate of $10^{-4}$, and a conditional dropout $p_{cond} = 20\%$, until the loss converged after $300,000$ epochs. 

\subsection{Quantitative Diagnostics}
\label{section:quant_methods}
Quantitative evaluation of the structures generated by the trained DDPM was performed using both material science domain metrics and machine vision metrics. Crystallographic texture was encoded using the previously described GSH expansion, and morphology with two-point correlation functions. Because these metrics are high dimensional representations, each was independently reduced to two- and three-dimensional representations using Principal Component Analysis (PCA). For texture space, the first two principal components (GSH PC-1 and GSH PC-2) captured $17.2\%$ and $16.8\%$ respectively of the total variance. For Spatial sStatistics (SS) space, SS PC-1, SS PC-2, and SS PC-3, captured $77.6\%$ and $9.7\%$, and $6.8\%$ respectively. Overlap between the generated and target microstructural descriptors in their respective PCA-projected spaces indicates high statistical similarity. 

Unlike the previous metrics, the property information is already a two-dimensional representation, therefore, no additional projection was required. However, in order to assess the consistency of property conditions between the target and generated microstructures, an estimate of the mechanical properties of the generated samples is required. Instead of simulating the mechanical response of each generated sample using CPFEM, we trained a surrogate CNN-to-MLP network that predicts the yield strength and the uniaxial elastic modulus from Euler angles. Lastly, process conditioning fidelity was quantified class-wise with the Fréchet Inception Distance (FID) metric \cite{heusel2017gans}. FID measures how similar a generated dataset is to a reference dataset by comparing the features extracted from both datasets using the pretrained Inception-V3 network, which is trained on natural images. More specifically, class-wise FID compares features from each process RVE independently, here,  \textit{fidelity} means similarity between the feature distribution of an RVE sample and the that of the corresponding process-conditioned sample. A lower FID score indicates a larger similarity between real and generated samples. Since FID is a statistic measure of distributional similarity, a large number of examples ($N=10000$) was used for each process condition.

\begin{figure*}[!t]
  \centering
  \includegraphics[width=1.0\linewidth, trim=0.0in 0.0in 0.0in 0.0in, clip]{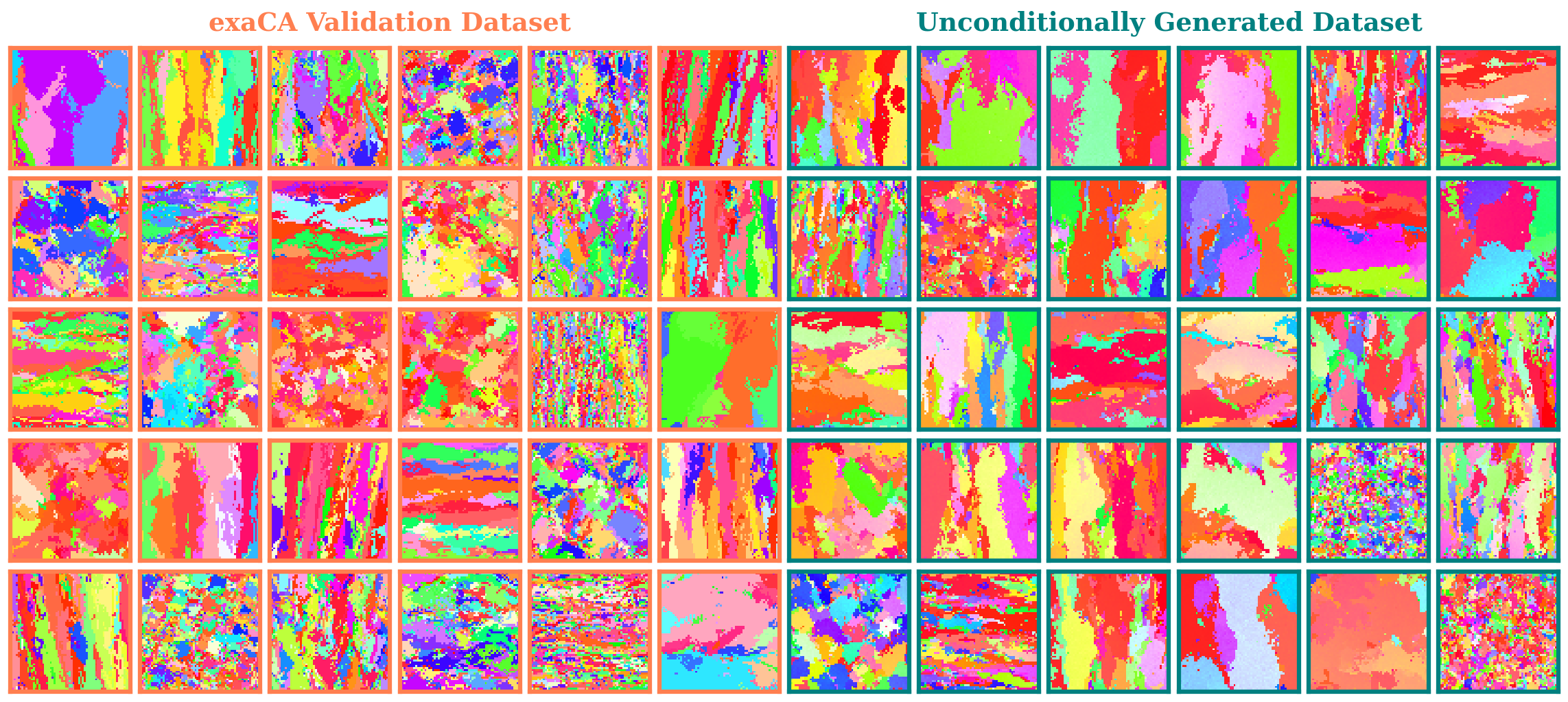}
  \caption{\textbf{Unconditional Generation Results.} Comparison of the validation dataset and the unconditionally generated micrographs as IPF-Z maps.}
  \label{fig:unconditional}
\end{figure*}

\section{Results}
In this section, we illustrate the generative performance of the DDPM in both unconditional and conditional generation and quantitatively evaluate sample fidelity and diversity by comparing microstructural descriptors between generated samples with those of the target data in our validation dataset. 

\subsection{Unconditional Generation}
A comparison between examples from the validation dataset and generated samples is shown in Fig. \ref{fig:unconditional}. Visually, texture and morphology distributions from the generated samples agree with the structures present in the validation dataset. The generated micrographs exhibit highly complicated spatial arrangements that cannot be adequately captured by traditional statistical sample descriptors and packing-based reconstruction methods. The DDPM correctly captured sharp orientation discontinuities at the grain boundaries, amorphous grain patches commonly observed in AM materials, fiber textures, and diverse grain sizes. The observed diversity of physical structures highlights the expressiveness and multimodal generative capability of the DDPM. Interestingly, generated samples contained internal grain misorientations that were not present in the training dataset. This artifact was also common across all conditioning examples.

\subsection{Conditional Generation}
Conditional generative quality was evaluated for each of the individual contextual options. For texture, morphology, and properties, the experiments were conducted using five unique targets. To ensure target diversity, we used a space-filling design algorithm \cite{joseph2015maximum} applied to the compressed representation associated with each contextual option, namely, a 2D texture space, a 3D spatial statistics space, and a 2D property space. We generated 15 micrographs per conditioning example, computed the corresponding conditioning metric, compressed the representation with PCA, and plotted the results together with the reference validation dataset in the same space. Additionally, we performed multiple sampling experiments to assess the effect of guidance weight $w$ over the final generative quality. These experiments were repeated for multiple conditioning strengths to asses the sensitivity of this hyperparameter.

\begin{figure*}[!t]
    \centering
    \begin{subfigure}[b]{0.48\textwidth}
        \centering
        \includegraphics[width=\textwidth]{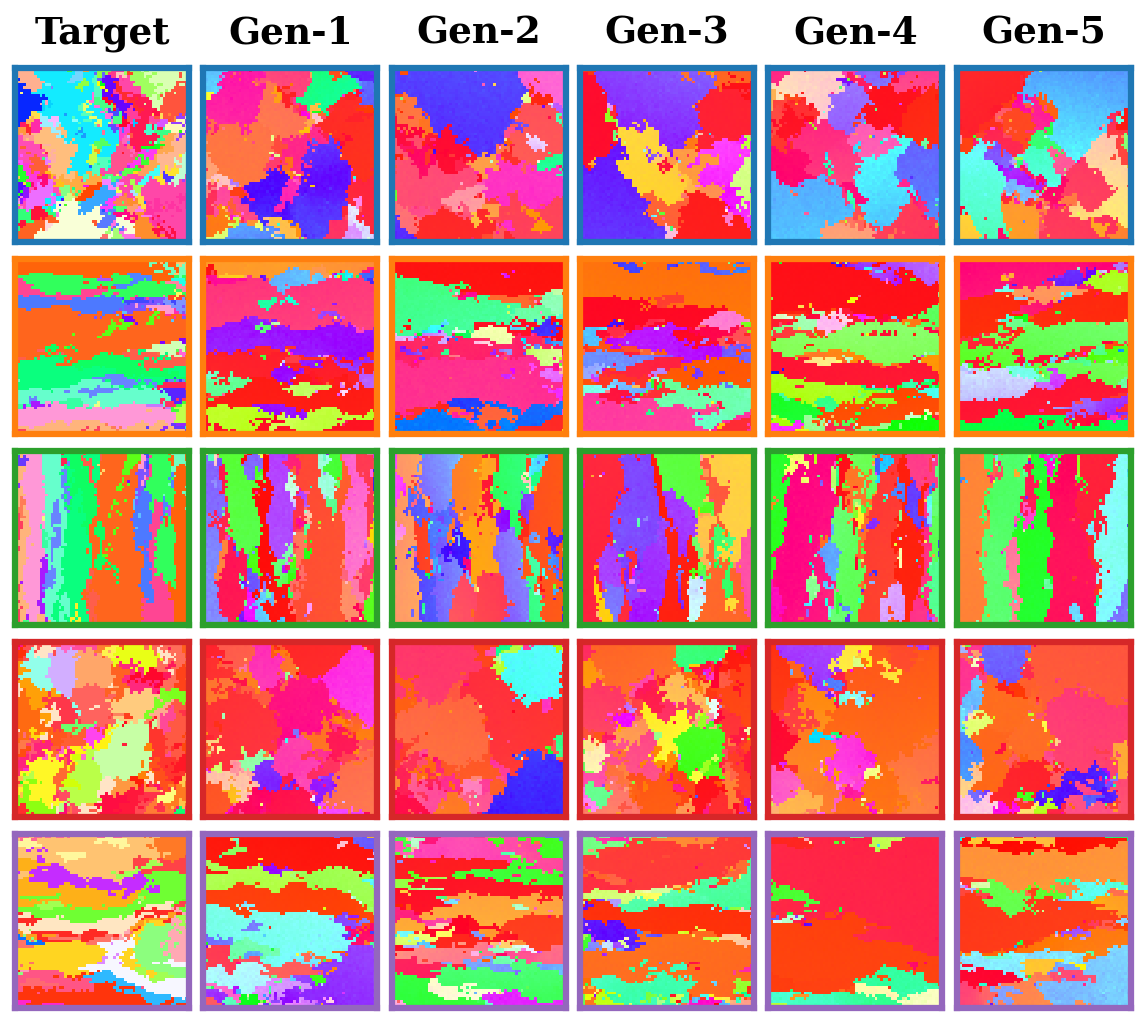}
        \caption{}
        \label{fig:results_morph_a}
    \end{subfigure}\hfill
    \begin{subfigure}[b]{0.48\textwidth}
        \flushleft
        \includegraphics[width=\textwidth]{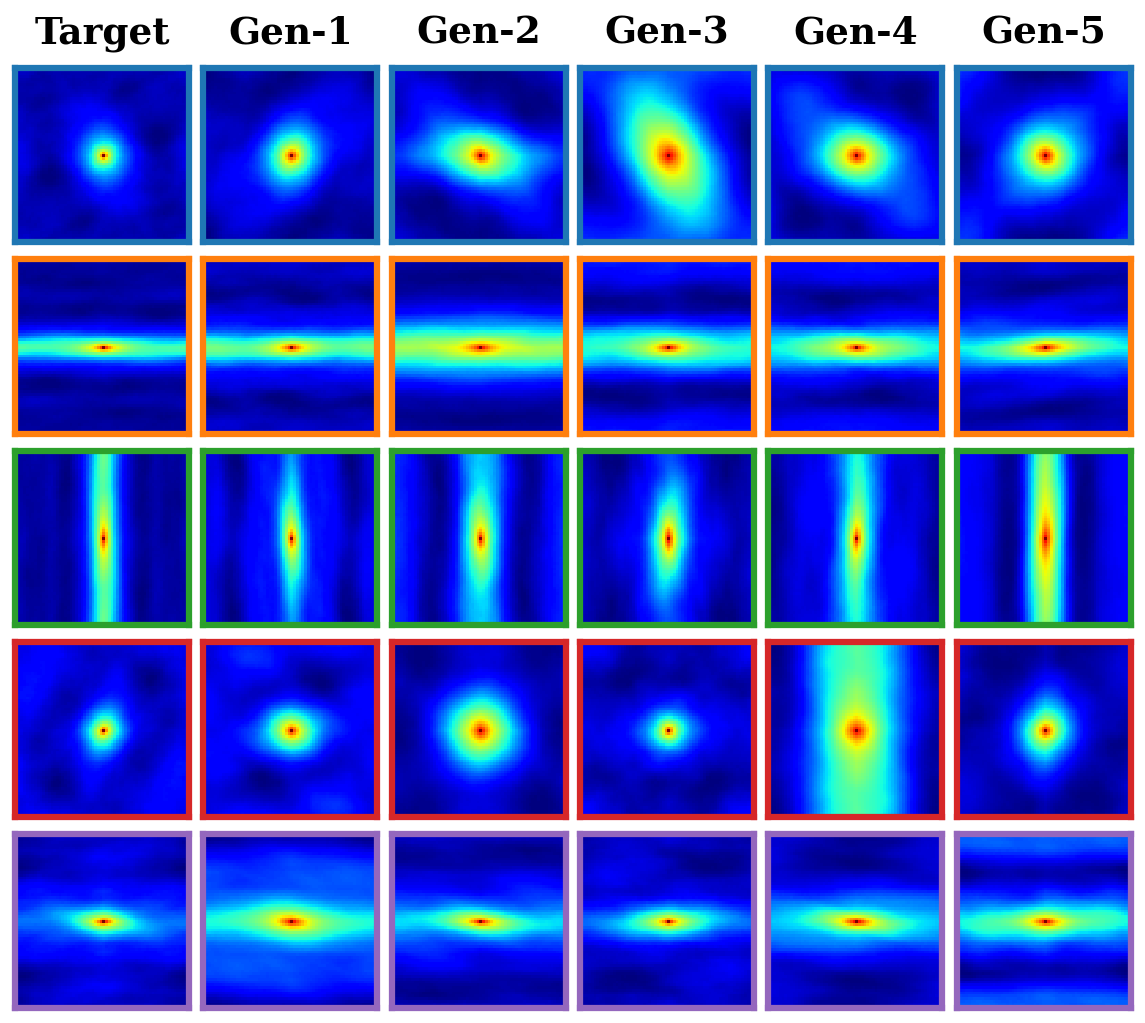}
        \caption{}
        \label{fig:results_morph_b}
    \end{subfigure}

    \vspace{0.5cm} 

    \begin{subfigure}[b]{0.48\textwidth}
        \centering
        \includegraphics[width=\textwidth]{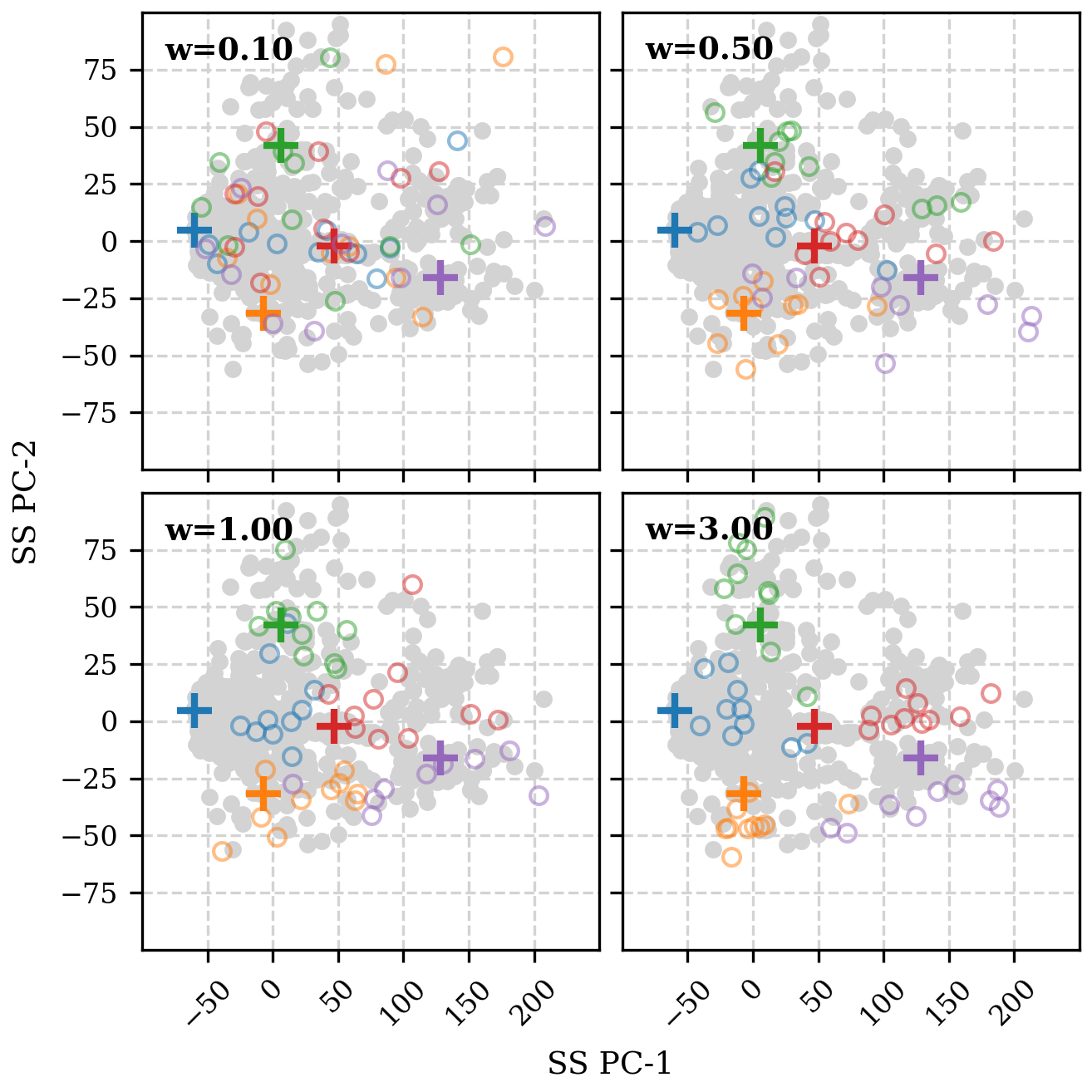}
        \caption{}
        \label{fig:results_morph_c}
    \end{subfigure}\hfill
    \begin{subfigure}[b]{0.48\textwidth}
        \centering
        \includegraphics[width=\textwidth]{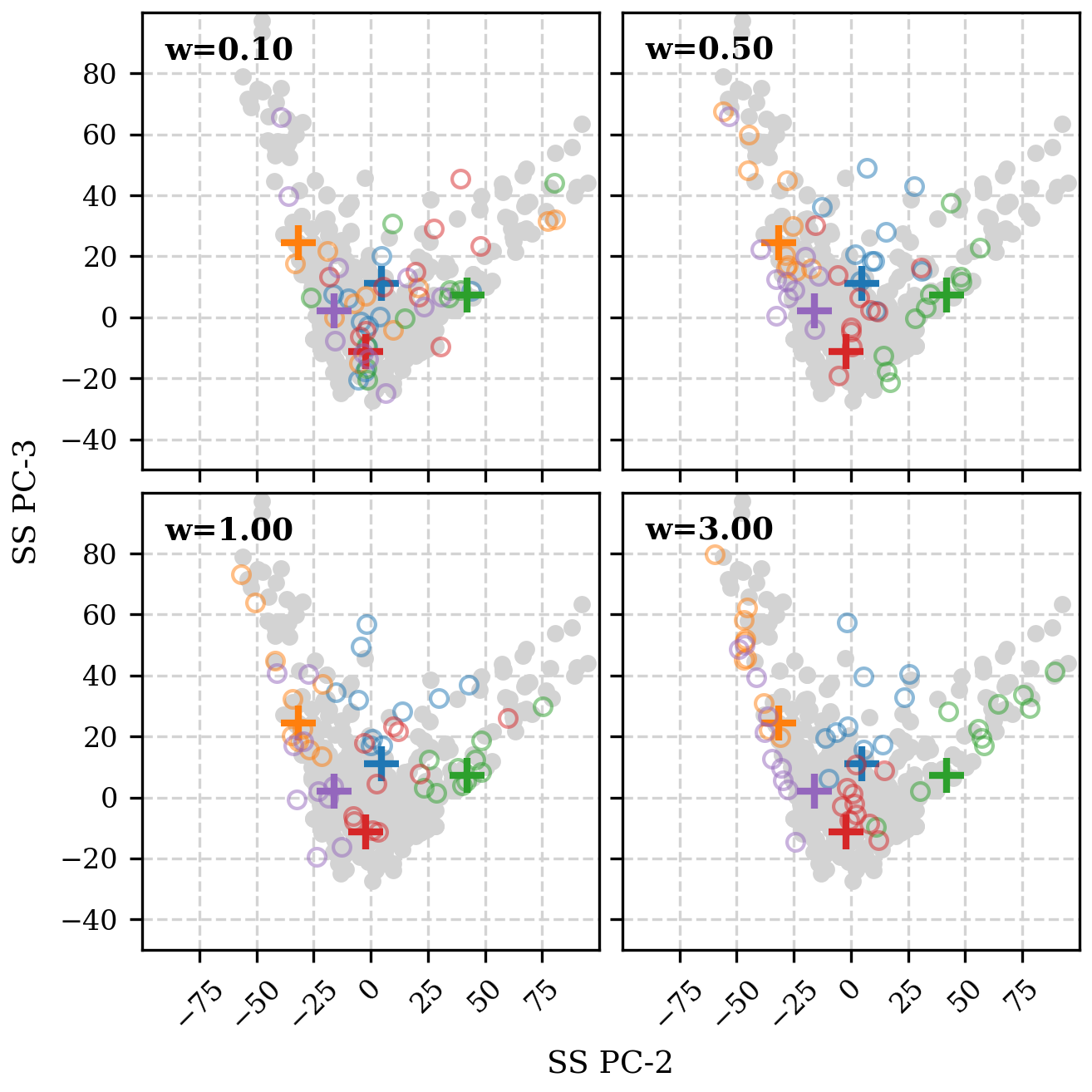}
        \caption{}
        \label{fig:results_morph_d}
    \end{subfigure}
    
    \caption{\textbf{Morphology-conditioning Results}. (a) IPF-Z maps generated with $w=1.0$ and $N=1$. Each row is a unique example where \textit{Target} is a data point from the validation dataset, \textit{Gen 1-5} are generated samples conditioned on the \textit{target} morphology. (b) Autocorrelation maps corresponding to each IPF map. (c-d) Quantitative diagnostic plots of generated samples in spacial statistics (SS) PC-space}
    \label{fig:conditional_morph}
\end{figure*}

\subsubsection{Morphology Conditioning}
The spatial structure of the generated samples in Fig. \ref{fig:conditional_morph} were consistent with those of the target microstructures. The DDPM is expressive enough to distinguish whether the sampled structure exhibits an equiaxed, horizontally columnar or vertically columnar microstructure. No mode collapse was observed in the generated examples, as the generated grains remained diverse. 

Quantitatively, we observed that $w$ impacted the variance of the generated sample statistics. However, while the morphologies are visually consistent, in the PC-space diagnostic plots, samples are scattered around the target. Interestingly, despite PC-1 capturing a substantial explained variance of $77.6\%$, the generated samples exhibited a significant dispersion. Each principal axis is expected to describe a unique morphological feature, see the Supplementary Material for a study about PCA applied to spatial statistics. We concluded that the variance associated with long-range correlations outweighs the subtle variations present in short-range correlations, leading PC-1 to capture global features such as texture rather than morphology. Hence, while our computation of spatial statistics omits some information about texture, such as the specific fiber type and orientation, it preserves whether a material is textured (giving rise to long-range correlations) or not. The scatter in the SS PC diagnostic plots arises from the omission of explicit crystallographic texture information. Thus, it is not an artifact of the DDPM but rather a consequence of the intrinsic behavior of our morphological statistics, which also encode \textit{some} texture information.

\begin{figure}[!t]
    \centering
    \begin{subfigure}[b]{0.48\textwidth}
        \centering
        \includegraphics[width=\textwidth]{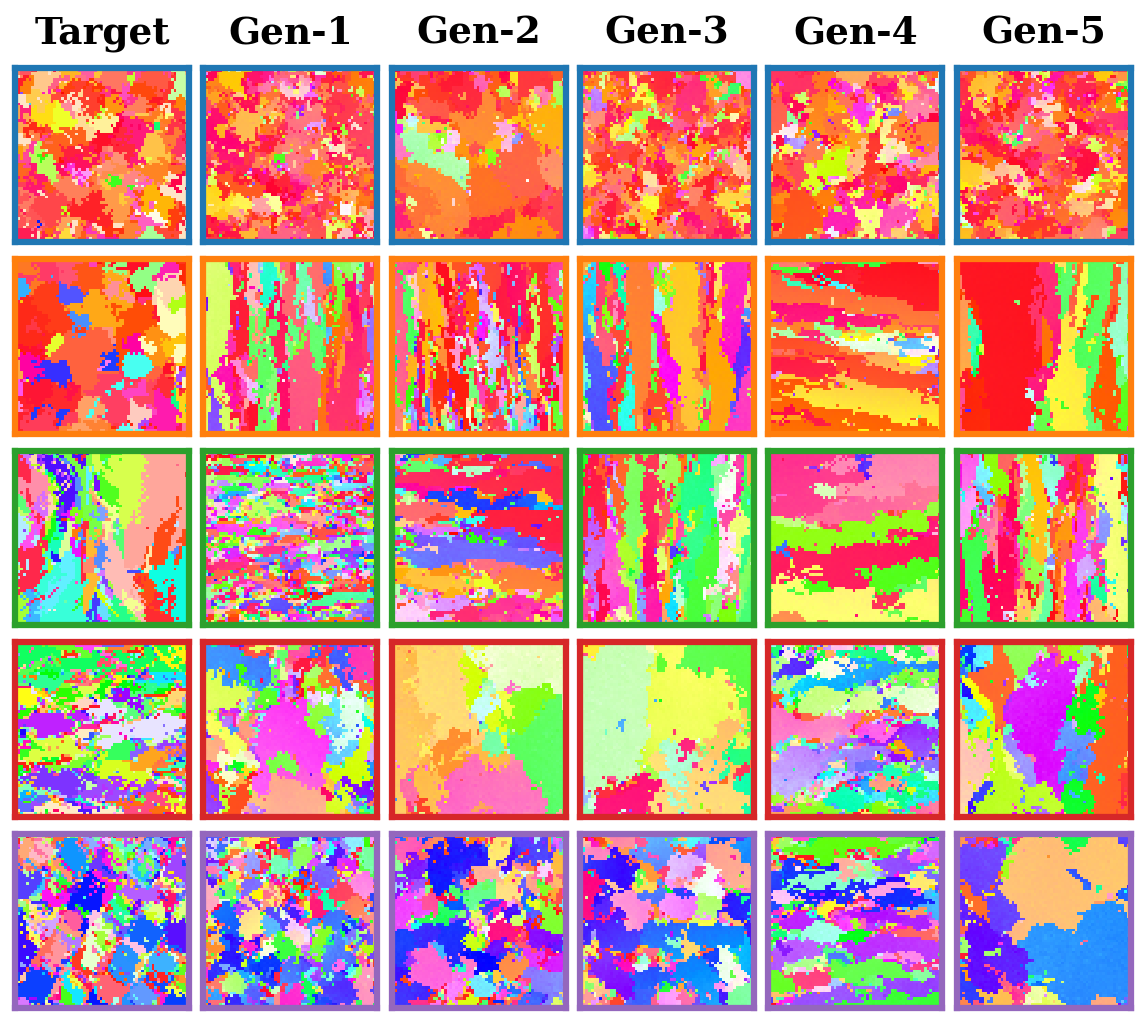}
        \caption{}
        \label{fig:ipf_prop_a}
    \end{subfigure}

    \vspace{0.5cm} 
    \begin{subfigure}[b]{0.48\textwidth}
        \flushleft
        \includegraphics[width=\textwidth]{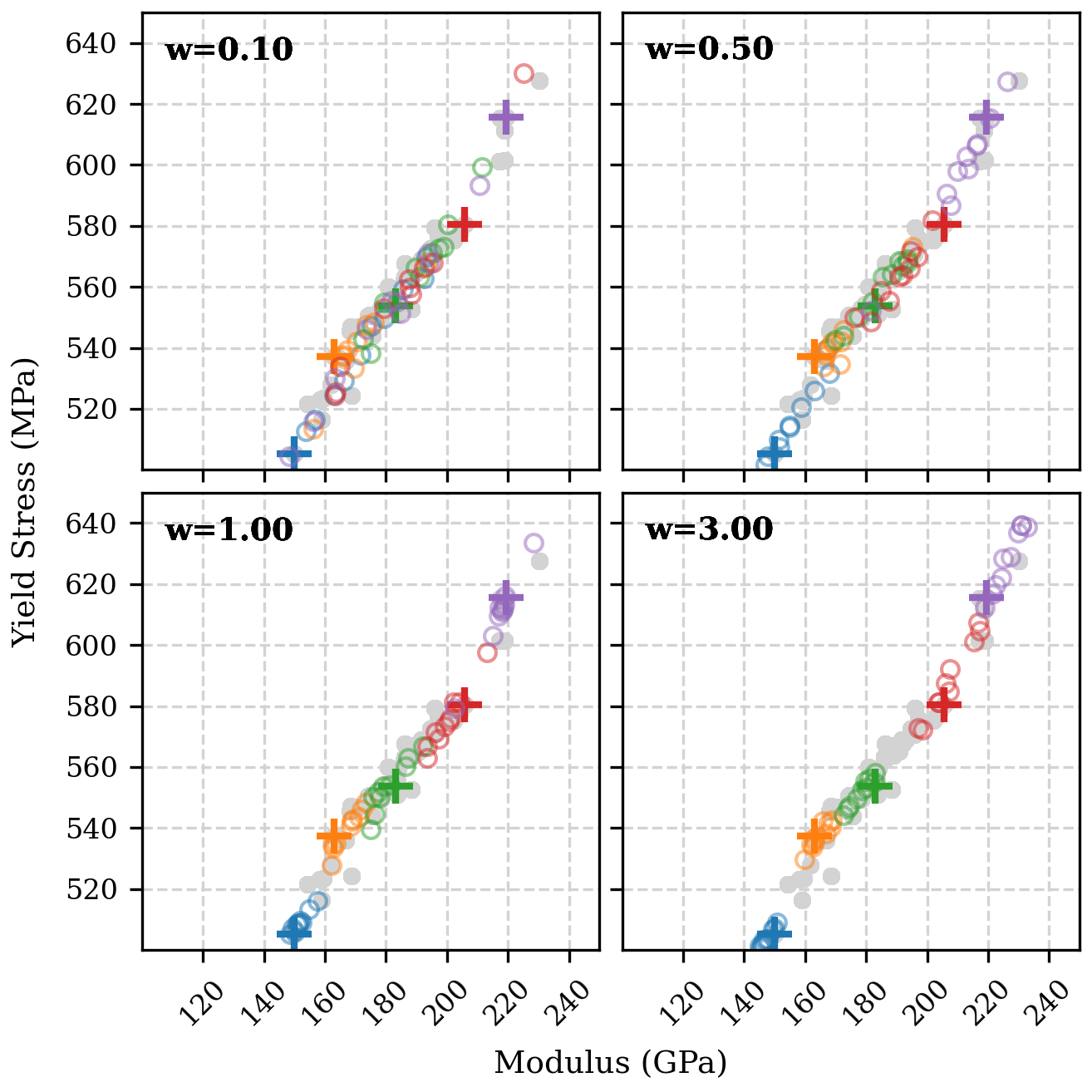}
        \caption{}
        \label{fig:ipf_prop_b}
    \end{subfigure}

    \caption{\textbf{Property-conditioning Results}. (a) IPF-Z maps generated with $w=1.0$ and $N=1$. Each row is a unique example where \textit{Target} is a data point from the validation dataset, \textit{Gen 1-5} are generated examples conditioned on the \textit{target} property label. (b) Quantitative diagnostic plots of the generated samples in property space}
    \label{fig:conditional_property}
\end{figure}

\subsubsection{Property Conditioning}
As previously described, the CPFEM simulations were performed on $64 \times 64 \times 64$ blocks, with the uniaxial loading applied along the Z-axis as shown in Fig. \ref{fig:data_example}. The training dataset consists of  XY- slices extracted from these 3D blocks. The uniaxial elastic modulus and yield strength for each XY-slice were obtained from the corresponding subsampled $64^3$ 3D DAMASK simulation. The IPFs of the property-conditioned generated micrographs are shown in Fig. \ref{fig:conditional_property}. All IPFs exhibit complex grain distributions that closely resembled the \textit{exaCA} dataset. Arranging the IPFs from top to bottom in ascending order according to their mechanical properties reveals a transition from $<001>$ to $<111>$ texture fiber. Fundamentally for FCC materials loaded along the fiber axis, a $<001>$ fiber texture is generally more compliant and weaker than a $<111>$ fiber \cite{thapliyal2024considering}, and CPFEM simulations inherently capture this physical relationship. The clear texture transition observed in the generated micrographs is particularly impressive, as it suggests that the DDPM successfully learned this fundamental structure-property relationship using only two property parameters. More importantly, this result highlights the remarkable potential of DDPMs to solve highly complex inverse problems such as property-to-structure mapping.

Visually, generated micrographs exhibited different morphologies while preserving a texture similar to that of the target micrograph, indicating that property conditioning prioritized matching the target texture rather than its morphology. This can be attributed to the augmentation scheme applied to the dataset. Because the loading direction is out of plane and augmented samples were generated by rotations about this axis, the property conditioning is insensitive to the specific in-plane morphological orientation of grains. Moreover, because the CPFEM model does not explicitly account for grain-size sensitivity, this information is not encoded in the data, and the simulated properties therefore depend primarily on texture. 

The property data corresponding to the generated examples correspond to the predictions of the previously described surrogate structure-property network. Again, it is visible that the conditioning strength $w$ influences how strongly the generative process adheres to conditioning information. However, this comes at the cost of reduced generative diversity, as reflected in the decreasing variability of the generated micrographs around the target statistic. While some of the variability may be attributed to the predictive error from the surrogate model, there is a clear agreement nonetheless between the target and conditionally generated micrographs.

\begin{figure*}[!t]
    \centering
    \begin{subfigure}[b]{0.48\textwidth}
        \centering
        \includegraphics[width=\textwidth]{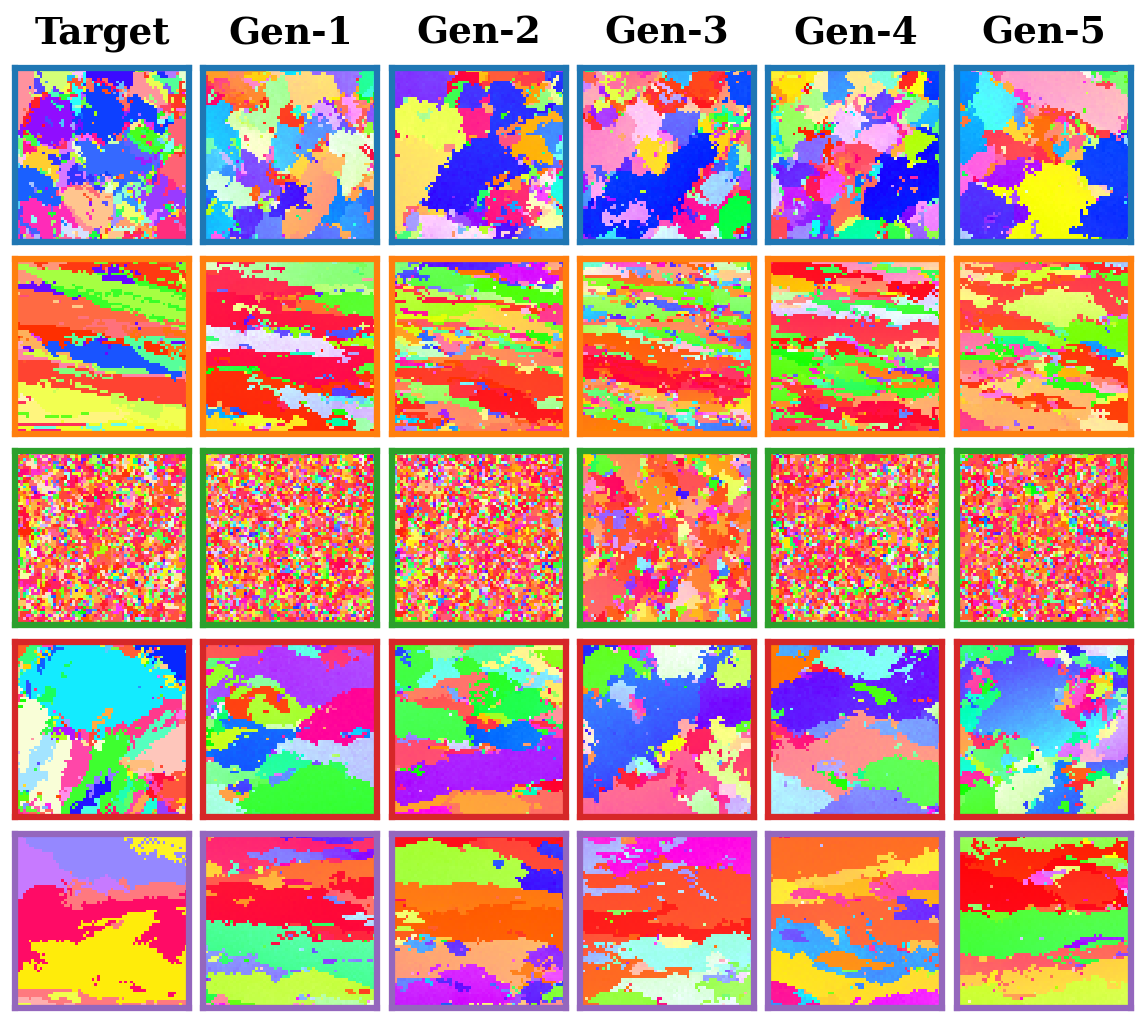}
        \caption{}
        \label{fig:ipf_texture_a}
    \end{subfigure}\hfill
    \begin{subfigure}[b]{0.48\textwidth}
        \flushleft
        \includegraphics[width=\textwidth]{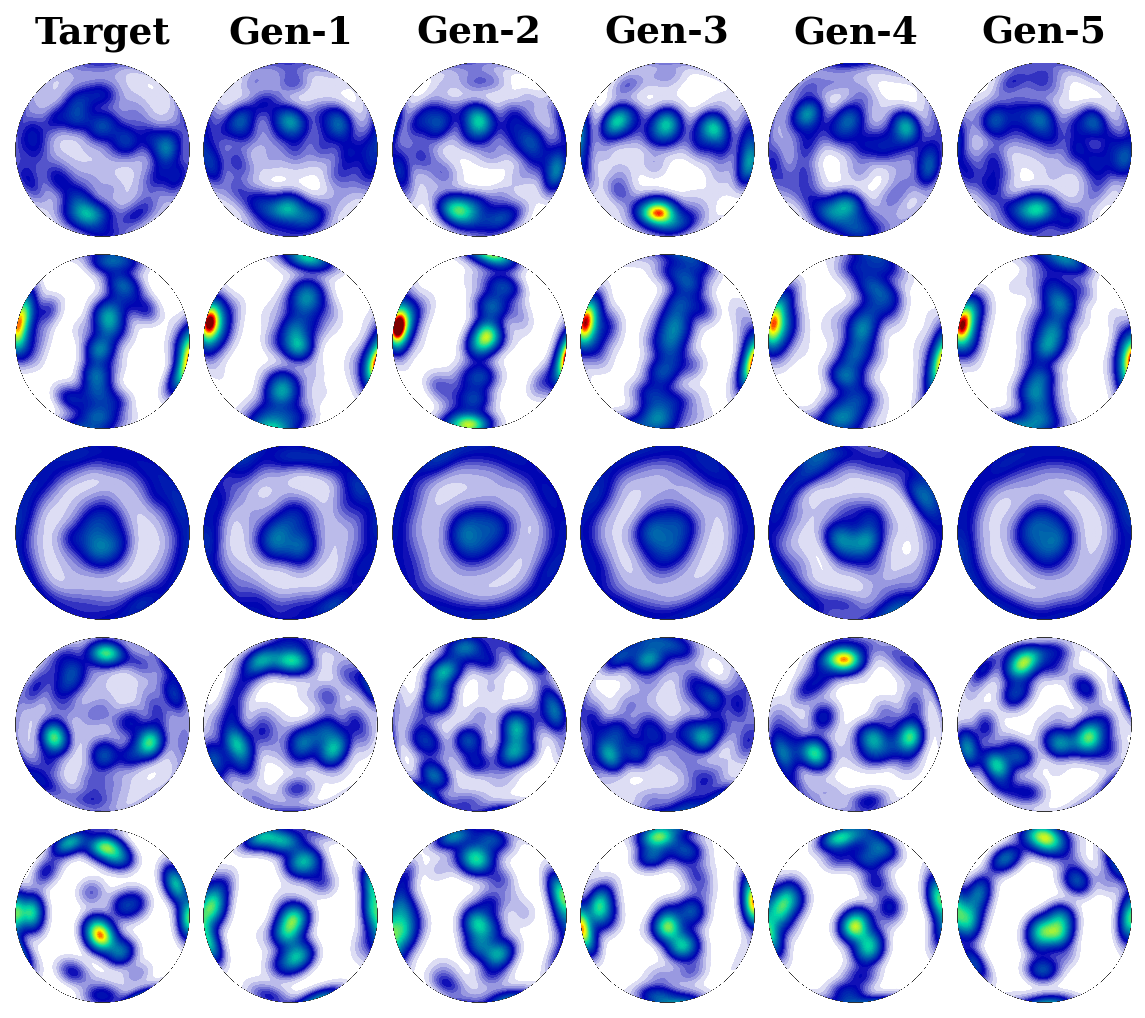}
        \caption{}
        \label{fig:ipf_texture_b}
    \end{subfigure}

    \caption{\textbf{Texture-conditioning Results}. (a) IPF-Z maps generated with to $w=1.0$ and $N=1$. Each row is a unique example where \textit{Target} is a data point from the validation dataset, \textit{Gen 1-5} are generated micrographs conditioned on the \textit{target} texture. (b) Corresponding $<001>$ pole figures.}
    \label{fig:conditional_texture}
\end{figure*}

\begin{figure}[!tbh]
  \centering

  \includegraphics[width=1.0\linewidth, trim=0.0in 0.0in 0.0in 0.0in, clip]{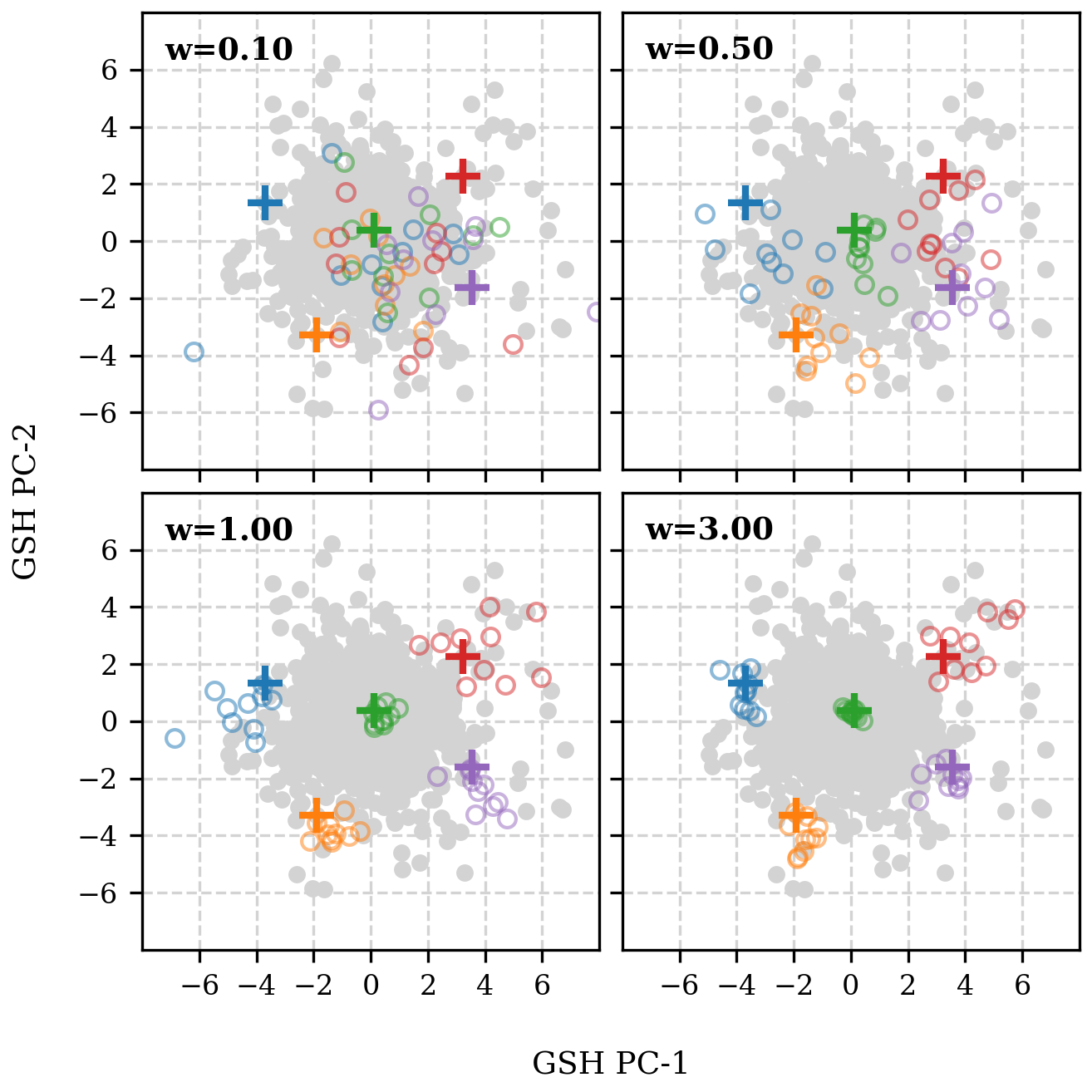}
  \caption{\textbf{Texture-conditioning Diagnostics.} Visualization of the generated samples from Fig. \ref{fig:conditional_texture} in the generalized spherical harmonics (GSH) PC-space}
  \label{fig:conditional texture_quant}
\end{figure}

\subsubsection{Texture Conditioning}

Texture-conditioned examples as seen in Fig \ref{fig:conditional_texture} and \ref{fig:conditional texture_quant} reproduce the orientation distribution from the target micrographs across many cases, including both weakly and strongly textured examples. Pole figures comparing generated and target examples also confirm that the generated samples exhibit similar ODFs and that the GSH-based conditioning scheme is effective. Despite the lack of explicit morphological information, the grain shapes in the generated micrographs are visually consistent with the targets. This reflects the inherent correlation between certain textures and specific morphologies, for example, fiber textures are generally associated with columnar grains. These findings highlight the ability of the conditioned DDPM in capturing deep nonlinear relationships among complicated physical features. Furthermore, they provide evidence that the DDPM and the self-supervised training strategy are sufficiently effective for the learned model to capture these embedded structure-structure cross-correlations.

The statistical fidelity of an ODF depends directly on the number of grains captured by the EBSD scan. When a map samples fewer grains, the resulting ODF tends to exhibit discrete orientation spikes that do not adequately represent the overall texture. Despite the difficulty to model such sparse distributions, the DDPM successfully reconstructs ODFs from coarse grained micrographs. For instance, the generated samples ODFs from the target in the fourth row remain similar to that of the target, despite the presence of large grains and the limited sample statistics.

The diagnostic plots in GSH-PC space illustrate the influence of $w$ on conditioning. Clustering around the target provides evidence of texture consistency. Notably, the dispersion of the generated micrographs differs across targets, for example, compare the red and green targets. These targets differ in grain sizes, with the red target exhibiting fine grains and the green target exhibiting coarse grains. Because texture is inherently a statistical quantity, cases in which fewer grains captured in the micrographs are expected to exhibit greater variance. These findings demonstrate the impressive capability of DDPMs to capture the statistical relationship between grain sampling and texture variability.

\begin{figure}[!t]
  \centering
  \includegraphics[width=1.0\linewidth, trim=0.0in 0.0in 0.0in 0.0in, clip]{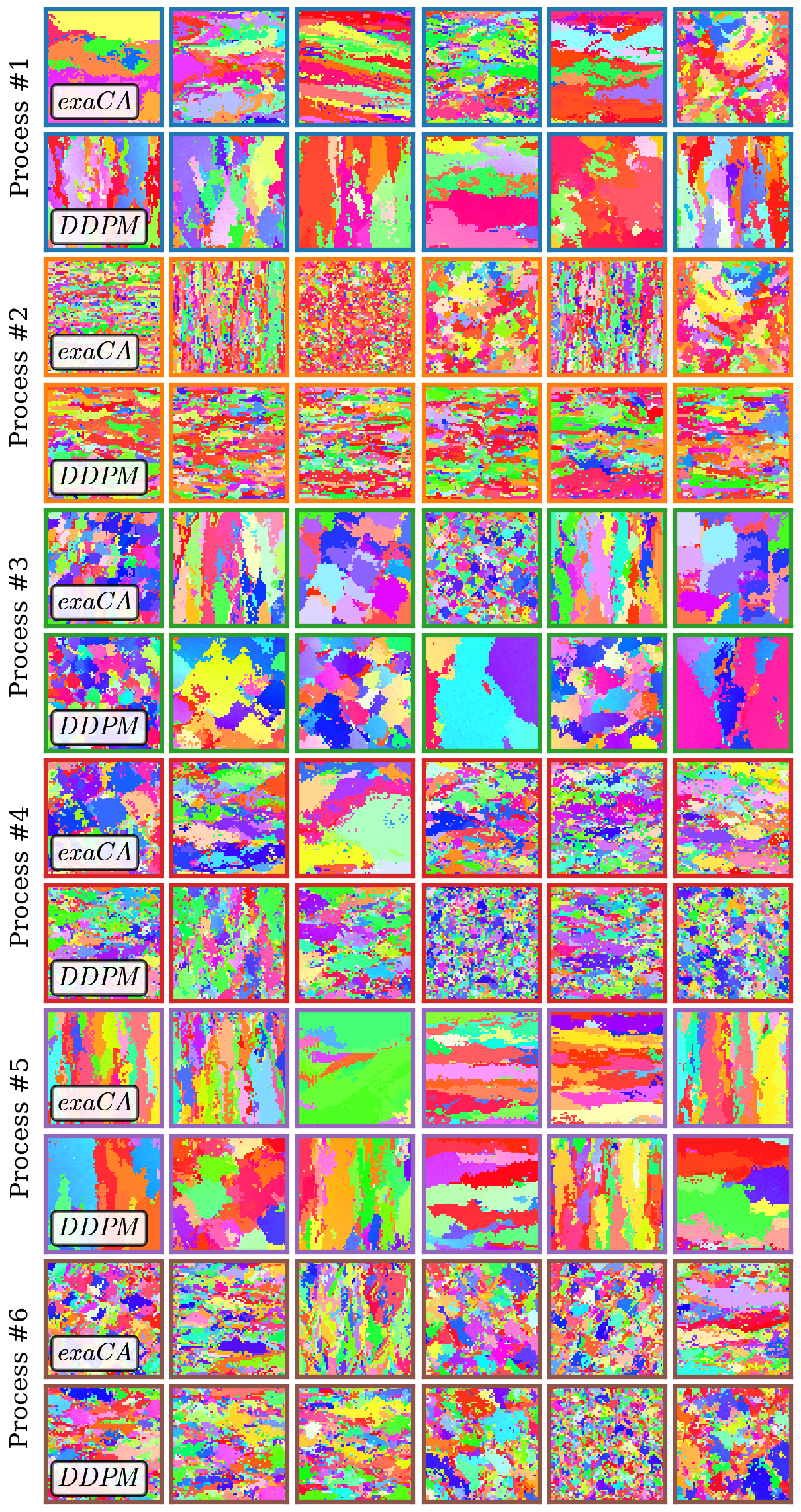}
  \caption{\textbf{Process-conditioning Qualitative Results.} IPF-Z maps generated with $w=1.0$ and $N=1$. \textit{exaCA} are samples from the validation dataset, \textit{DDPM} are samples generated with its respective process class labeled on the left.}
  \label{fig:conditional process}
\end{figure}

\begin{figure}[tbp]
  \centering
  \includegraphics[width=1.0\linewidth, trim=0.0in 0.0in 0.0in 0.0in, clip]{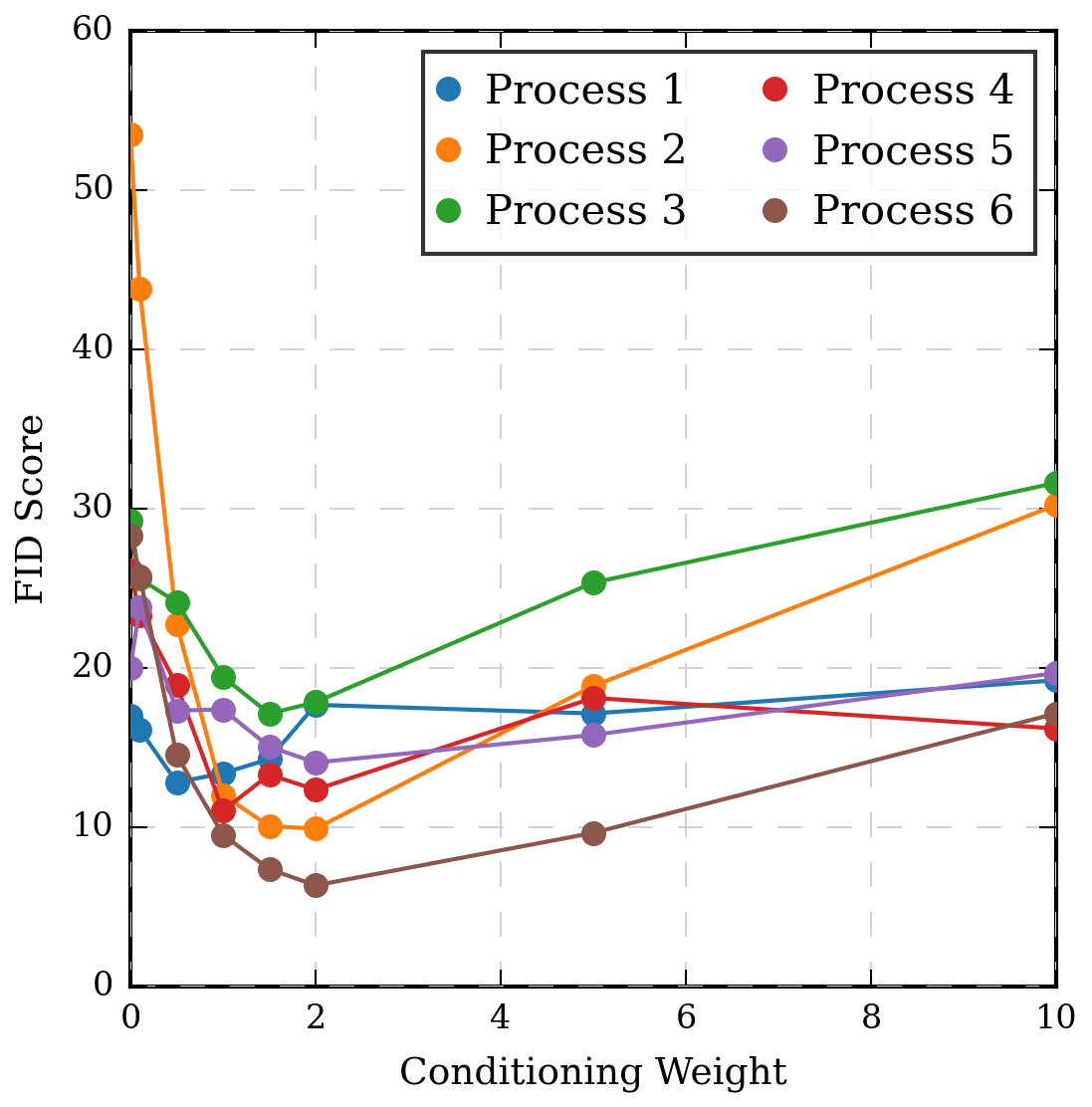}
  \caption{\textbf{FID Score Evaluation} FID score comparison of process-conditioned samples across multiple conditioning weights.}
  \label{fig:process_fid}
\end{figure}

\subsubsection{Process Conditioning}
Process is a discrete class representing the six unique sets of parameters used to generate the simulated the RVEs. Hence, this categorical feature was encoded within the conditioning vector using one-hot encoding.  Note that, in the present setup, the process condition was divided into six classes due to the limited dataset size. However, more generally, process-conditioning information could also be encoded explicitly using process parameters such as power, speed and hatch spacing.

The results of the process-conditioned generation are shown in \ref{fig:conditional process}. Visually, the IPFs of the generated micrographs closely match the overall texture, repeated grain patterns, and distinct columnar or equiaxed grained structures that differentiate each of the \textit{exaCA} RVEs. These results suggest that the DDPM effectively captured the underlying distribution associated with each of the six process parameter sets and, more generally, it may be capable of learning meaningful process-structure relationships relevant to inverse design. 

In order to evaluate process similarity quantitatively, we conducted an FID analysis on each process class across multiple guidance weights $w$.  The FID score results, shown in \ref{fig:process_fid}, exhibit a clear decrease as as $w$ increases, indicating a stronger feature agreement between the \textit{exaCA} dataset and the generated micrograph. This trend is consistent with our previous observations across all other conditioning cases. Interestingly, for all process classes, there is a turning point at around $w\approx 2$, beyond which the FID score begins to increase again, suggesting that overemphasizing $w$ comes at the expense of visual fidelity.  The same overemphasis on conditioning behavior may extend to all other conditions, which highlights the importance of hyperparameter tuning to maximize the generative quality associated with each independent condition.

\begin{figure}[!t]
    \centering
    \begin{subfigure}[b]{0.48\textwidth}
        \centering
        \includegraphics[width=\textwidth]{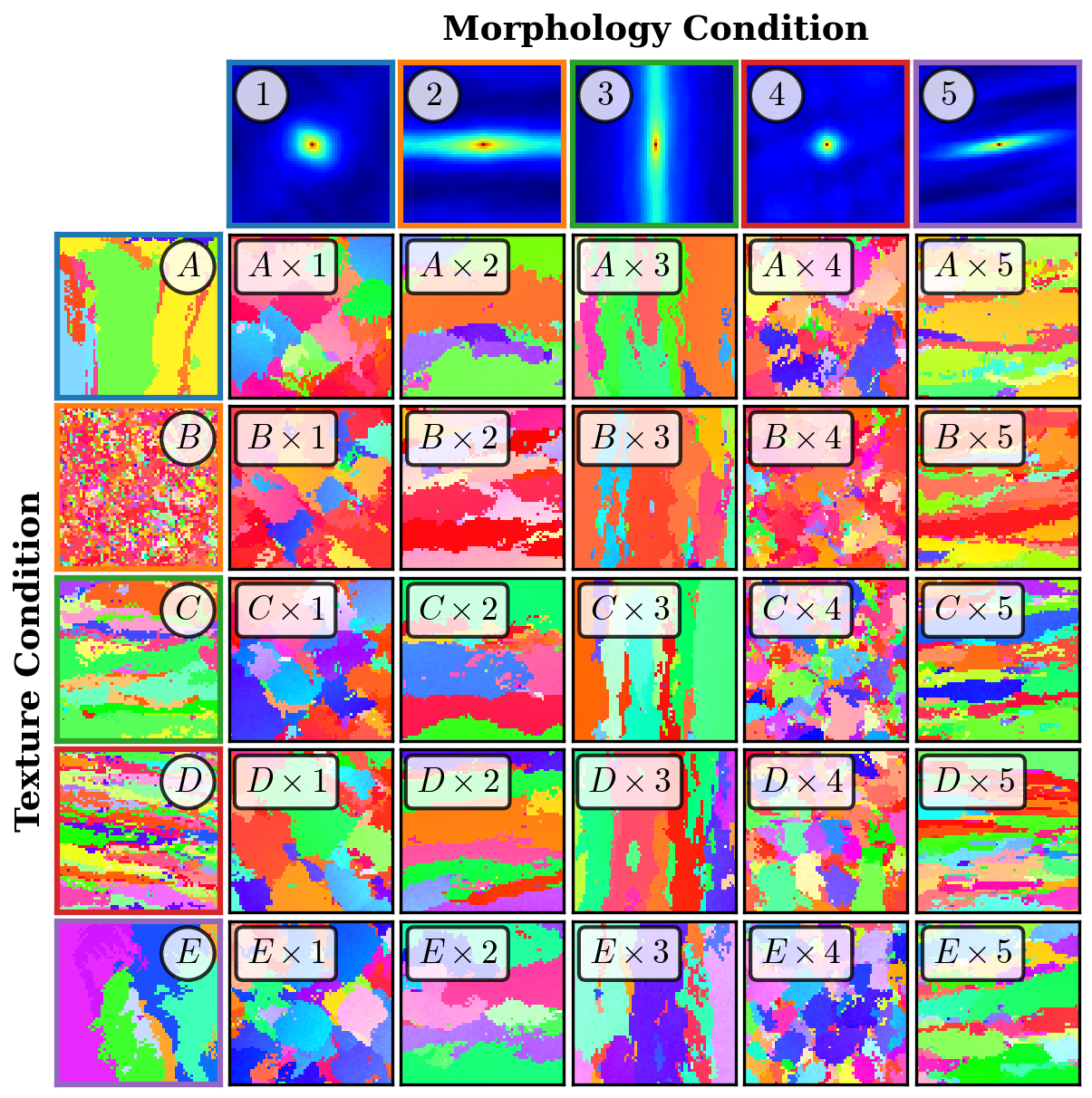}
        \caption{}
        \label{fig:ipf_texture_morph_a}
    \end{subfigure}

    \vspace{0.5cm} 
    \begin{subfigure}[b]{0.48\textwidth}
        \flushleft
        \includegraphics[width=\textwidth]{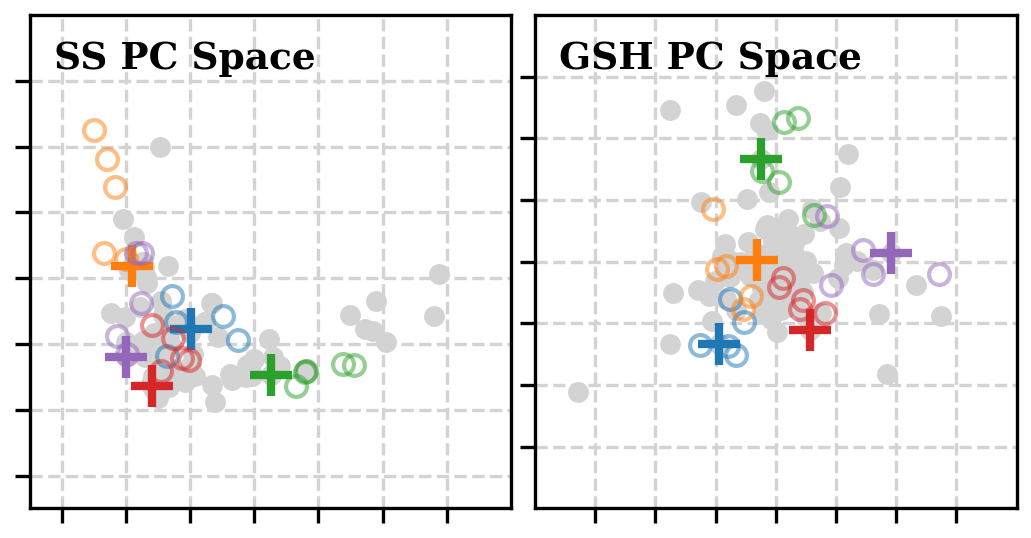}
        \caption{}
        \label{fig:ipf_texture_morph_b}
    \end{subfigure}

    \caption{\textbf{Texture and Morphology Cross-conditioning Results.} (a) IPF-Z maps of generated and target samples, top row corresponds to the target morphology condition, left column corresponds to the target texture condition. The letter and number at the top left corner of each sample describe the pair of texture and morphology condition used. Each sample was generated from a cross-conditioning of the respective target morphology and texture condition. (b) Spatial statistics PC-2 and PC-3 diagnostic plot (left), and texture generalized spherical harmonics PC-1 and PC-2 diagnostic plot (right) }
    \label{fig:cross-conditioning_texture_morph}
\end{figure}

\begin{figure}[!t]
    \centering
    \begin{subfigure}[b]{0.48\textwidth}
        \centering
        \includegraphics[width=\textwidth]{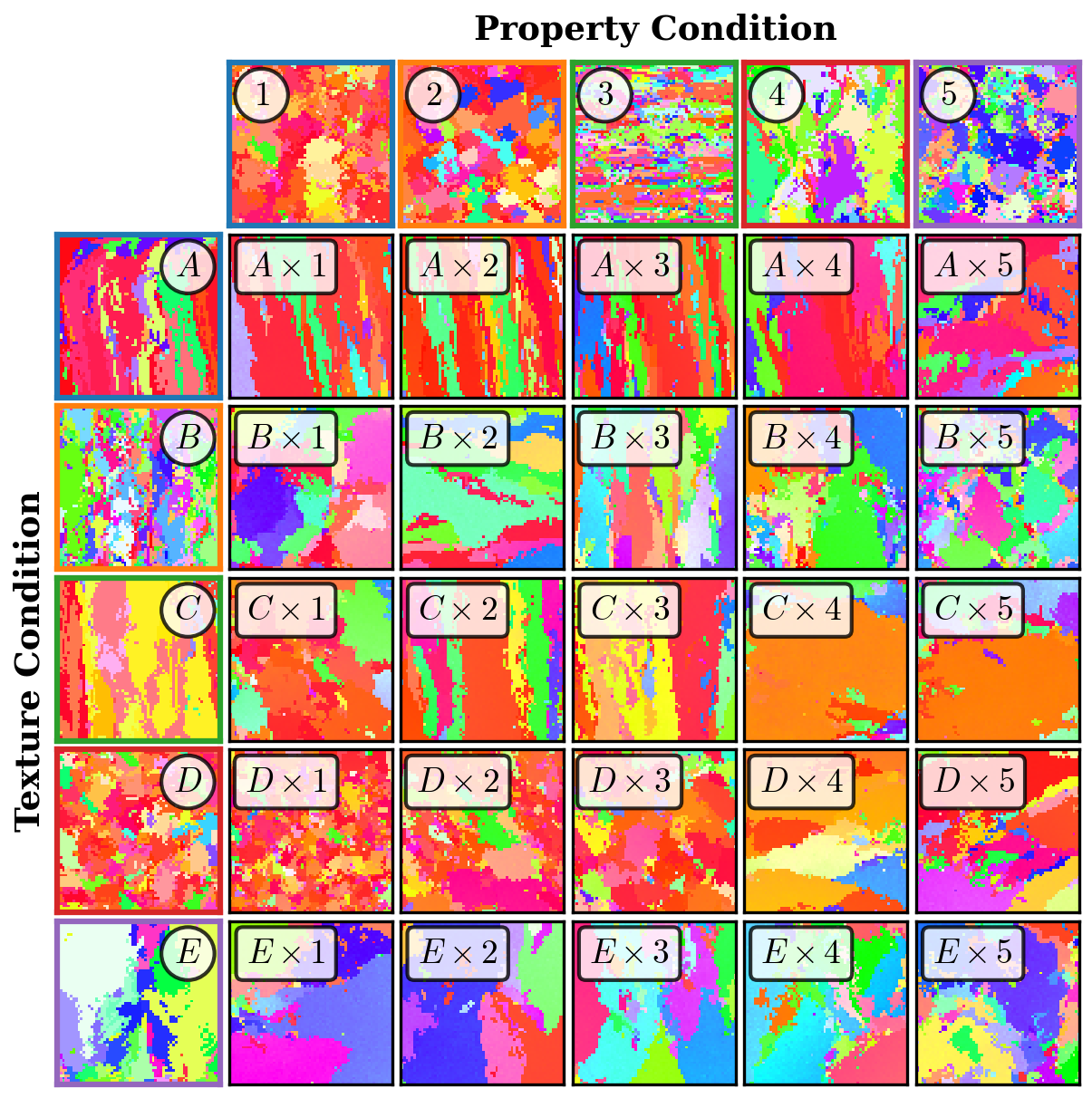}
        \caption{}
        \label{fig:ipf_texture_prop_a}
    \end{subfigure}

    \vspace{0.5cm} 
    \begin{subfigure}[b]{0.48\textwidth}
        \flushleft
        \includegraphics[width=\textwidth]{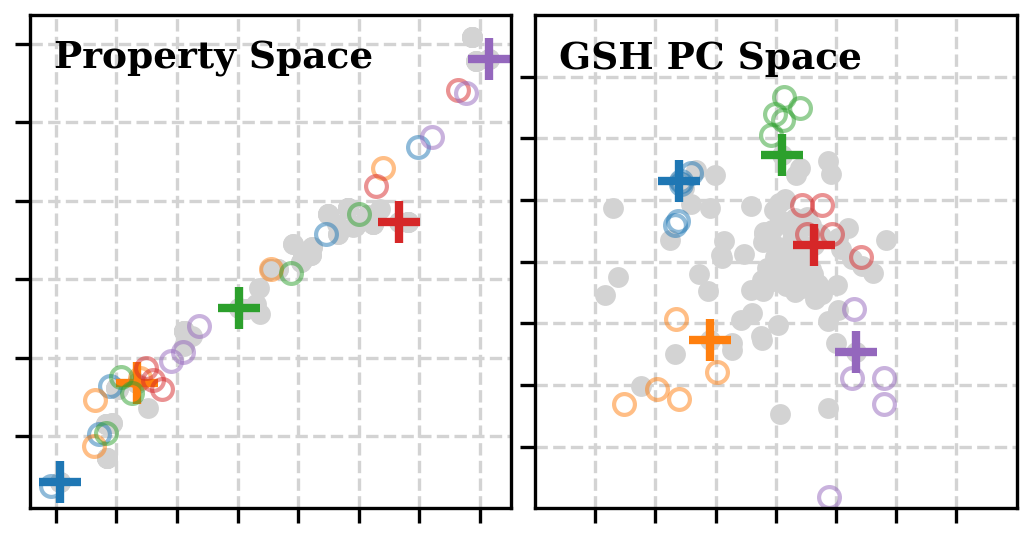}
        \caption{}
        \label{fig:ipf_texture_prop_b}
    \end{subfigure}

    \caption{\textbf{Texture and Property Cross-conditioning Results}.  (a) IPF-Z maps of generated and target samples, top row corresponds to the target property condition, left column corresponds to the target texture condition. The letter and number at the top left corner of each sample describe the pair of texture and morphology condition used. Each sample was generated from a dual-conditioning of the respective target morphology and texture condition. (b) Samples plotted in property (Modulus vs Yield stress) space (left), and texture PCA space (right) }
    \label{fig:cross-conditioning_texture_prop}
\end{figure}

\subsection{Cross-conditioning Generation}
Additional experiments were conducted to evaluate the ability of the DDPM to perform cross-conditioning generation to synthesize novel microstructures, such as a unique combinations of texture and morphology not present in the training dataset. The experiments consisted of two conditioning scenarios: first, conditioning on texture and morphology, and second, conditioning on texture and mechanical properties. These scenarios were selected to evaluate the generative performance under both physically plausible and potentially incompatible combinations of conditions. 

In this first experiment, we asses if the DDPM can generate novel, previously unseen examples or if it is limited to the original distributions. Textures and morphologies are not strictly linked, as a wide range of textures can exist across different morphologies, therefore, combinations of texture and morphology conditions unseen by the model may still physically compatible. If the model accurately has accurately captured the underlying distribution of the training dataset, it may be capable of interpolating between conditioning information and generate truly novel and physically plausible examples. To test this we generated samples using all possible combinations of five unique morphology and texture conditions obtained from the validation dateset. The generated IPF maps and diagnostic plots are shown in Fig. \ref{fig:cross-conditioning_texture_morph} . Visually, the generated examples are clearly consistent with both the texture and morphology condition. This is further supported by the PC diagnostic plots, in which the generated samples cluster around the target microstructure in both the SS PC space and the GSH PC space. Interestingly, although the samples were generated with a guidance $w=3.0$, the clusters are less compact than those obtained by conditioning individually. This difference is especially evident for texture conditioning as observed in Fig. \ref{fig:conditional_texture}. Suggesting that the DDPM trades off some texture variance in order to satisfy the morphology condition. This behavior reflects an implicit regularization effect that biases generation towards microstructures that resemble those in the training dataset. This is an remarkable result, as the DDPM appears capable of generating genuinely new microstructures while being implicitly self-regularized to preserve physically realistic features. 

While certain textures can generally occur independently of the specific morphology, the mechanical properties predicted by CPFEM are physically correlated with the texture type as shown in Fig. \ref{fig:conditional process}.  In the training dataset, a microstructure with a $<001>$ fiber never exhibited a higher uniaxial elastic modulus and yield strength than a  $<111>$ fiber. In this second experiment, we investigate how the model behaves when it is asked to generate physically incompatible combinations of texture and mechanical properties. As in the previous experiment, samples were generated for all possible combinations of five unique property and texture conditions, with the results shown in Fig \ref{fig:cross-conditioning_texture_prop}. Although the textures from the generated micrographs agree with the target condition, there is no clear pattern with property conditions. Quantitatively, while the generated samples cluster near the target in GSH space, the property diagnostic plot show no evidence of agreement with the targets. These results suggest that when two conditions that are physically incompatible, the DDPM tends to \textit{prioritize} one condition over the other. This behavior further indicates that the DDPM is self-constrained by the learned property-structure relationships, leading it to prioritize targets that remain consistent with the underlying physics encoded in the training dataset. Additional cross-conditioning experiments are provided in the supplementary material. 

\section{Discussion}
\label{discussion}
Across all conditioning cases, we evaluated the fidelity of the generated micrographs both qualitatively, through visual inspection of IPF maps, and quantitatively, using diagnostic plots. However, across both conditional and unconditional cases, we consistently observed color gradients within grains in the IPFs map, indicative of internal grain misorientation, a feature not present in our training dataset. In this section, we seek to investigate the source of behind this artifact, as well as potential mitigation strategies. In general, the guidance weight $w$ influenced the agreement between the target and generated micrograph with respect to the conditioning information (texture, morphology, process, property). However, for process-conditioned examples, overemphasizing $w$ led to an increase in FID scores, meaning that the distinct features of each process class were degraded. We explore the mechanism in DDPMs with CFG that leads to this decline in generative fidelity and emphasize the importance of careful hyperparameter tuning. The cross-conditioning experiments demonstrated that the DDPM can generate micrographs not seen in the training dataset and exhibit conditioning \textit{prioritization} when two physically incompatible conditions are used during generation, particularly prioritizing texture over mechanical properties. Through a comparison of diffusion scores, we provide insights into the potential order of conditioning prioritization and relate it to the microstructure information inherently encoded in the descriptor. Lastly, we demonstrate a potential use of our DDPM model within an uncertainty quantification framework. 

\begin{figure}[tbp]
  \centering
  \includegraphics[width=1.0\linewidth, trim=0.0in 0.0in 0.0in 0.0in, clip]{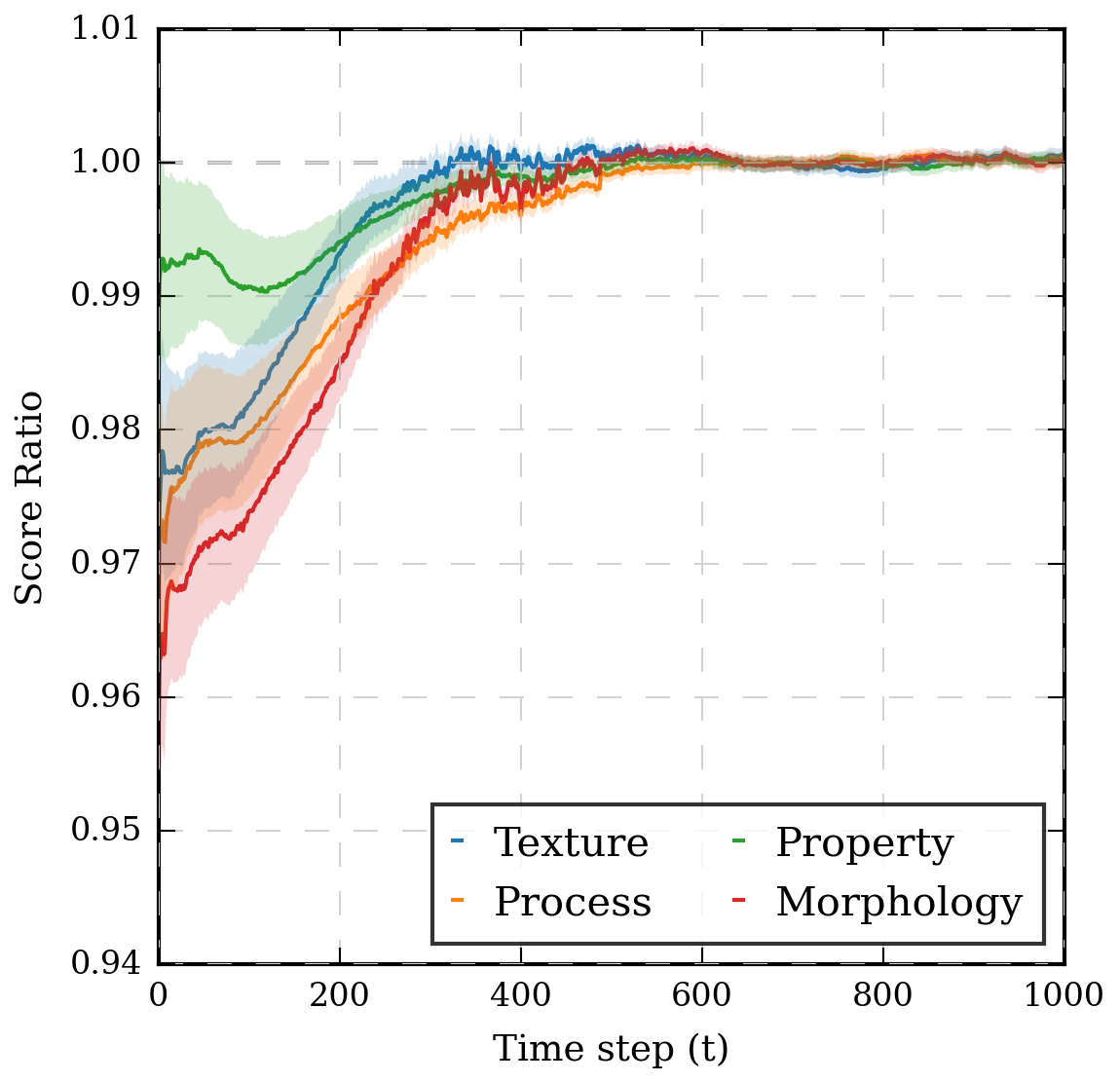}
  \caption{\textbf{Conditioning Preference Evaluation.} Diffusion score ratio evolution between conditional and unconditional generation. The score was estimated over $1000$ generated samples with $w=2.0$ and $N=1$ . A score ratio of one indicates the conditioning has no impact in the denoiser's effectiveness. A score ratio greater than one indicates that conditioning information results in a less effective denoising than in the unconditional case. A score ratio less than one indicates that conditioning improves the denoiser's effectiveness. }
  \label{fig:score_vs_t}
\end{figure}

\subsection{Condition Preference Evaluation}

Cross-conditioning with a physically incompatible combination of conditions resulted in a condition prioritization, as observed in Fig. \ref{fig:cross-conditioning_texture_prop}, where matching texture was prioritized over matching properties. To better understand this behavior, we quantitatively characterize the mechanism within the DDPM that determines which condition to prioritize. 

The reverse diffusion process is influenced by the amount of contextual information available about the target microstructure. Providing conditioning information, such as the presence of small grains or a sharp $<001>$ fiber texture, narrows the distribution of possible micrographs and, consequently, reduces ambiguity during sampling. From a noise-prediction perspective, this reduced ambiguity should make the noise in a corrupted micrograph easier to estimate, consequently, making the denoiser more effective. In order to quantify the extent to which each conditioning case reduces the ambiguity in noise predictions we use the \textit{diffusion score}. The diffusion score $\mathbf{s}_t(\mathbf{x})$, which is derived directly from Tweedie's formula \cite{luo2022understanding}, is a vector field that points towards the direction of steepest probability increase for a given $t$, and is defined as follows,

\begin{equation}
\begin{aligned}
\mathbf{s}_t(\mathbf{x}) = \nabla_\mathbf{x}\log{p_t(\mathbf{x}_t)} = -\frac{\bm{\epsilon}_{\theta}(\mathbf{x}_t,t)}{\sqrt{1-\bar{\alpha_t}}}
\label{eqn:diffusion_score},
\end{aligned}
\end{equation}

\noindent where $p_t(\mathbf{x})$ is the likelihood of observing $\mathbf{x}_t$ under the noisy data distribution given by the forward process at time step $t$. For an example visualizing the score function refer to the supplemental material. The score magnitude decreases as the sampled noisy state $x_t$ approaches the the local modes of $p(x_t)$. With conditioning information, the denoiser can guide the reverse process more directly toward these local modes. As a result, $x_t$ is expected to align with the target micrograph at $x_0$ earlier in the reverse process, which is reflected by a smaller score magnitude compared to the unconditional case. To quantify this effect, we computed the diffusion score ratio $\norm{\mathbf{s}_t(\mathbf{x},\mathbf{c})}^2/\norm{\mathbf{s}_t(\mathbf{x}, \varnothing)}^2$ , which provides an estimate of how conditioning improves the reverse process efficiency relative to unconditional generation. A score ratio of one indicates that the conditioning neither improves or degrades the effectiveness of the denoiser. A score ratio greater than one indicates that the introducing conditioning results in a less effective denoiser than in the unconditional case. Conversely, a score ratio less than one indicates that conditioning improves the effectiveness denoiser. 

Our results are shown in Fig. \ref{fig:score_vs_t} from which several conclusions can be drawn. Conditioning guidance began influencing diffusion at around $t=500$ for all conditioning cases.  This is because some of the important structure starts appearing half way through the diffusion process \cite{karras2022elucidating}. All conditions were descriptive enough to improve the sampling process, however, not all cases had the same degree of influence. On average, morphology conditioning had the most effective diffusion score, followed by process, texture, and lastly property conditioning. We hypothesize that the degree of influence is directly governed by how inherently informative the descriptor is about the overall sample structure. Spatial statistics quantify the local grain morphology of a microstructure through short-range correlations, while simultaneously capturing the presence of texture with long-range correlations (see supplemental section). Although the process conditioning contains information about the expected texture and morphology given an RVE, the morphological information may still vary due to the augmentation strategy, which includes random rotations and flips. For example, the same process label may correspond to longitudinal-, transverse- or equiaxed-grained morphologies, as well as horizontally or vertically flipped maps. While GSH coefficients only encode texture explicitly, certain textures happen to be somewhat correlated to specific grain morphologies in this specific AM dataset. Lastly, even though the yield strength and the uniaxial elastic modulus are both strongly influenced by texture and morphology, the uniaxial tensile condition along z-axis and the lack of grain-size sensitivity in the CPFEM simulations indicate that the property label is generally uninformative about the expected morphology. To summarize, the influence that conditioning may have on the reverse sampling process is fundamentally linked to how informative condition is about the overall microstructure. Conditioning information, such as morphology, process or texture, very narrowly constrain generation because they are explicitly operating on the structure itself. Information about properties is more indirectly related to the generated micrograph and fundamentally encompass a wider more diverse distribution, which therefore has less impact on denoiser effectiveness. With our diffusion score ratio study, we conclude that, when presented with two incompatible conditions, the DDPM prioritizes the condition associated with he most effective reverse diffusion process, therefore obeying the texture condition over the property condition. 

\begin{figure}[tbp]
  \centering
  \includegraphics[width=1.0\linewidth, trim=0.0in 0.0in 0.0in 0.0in, clip]{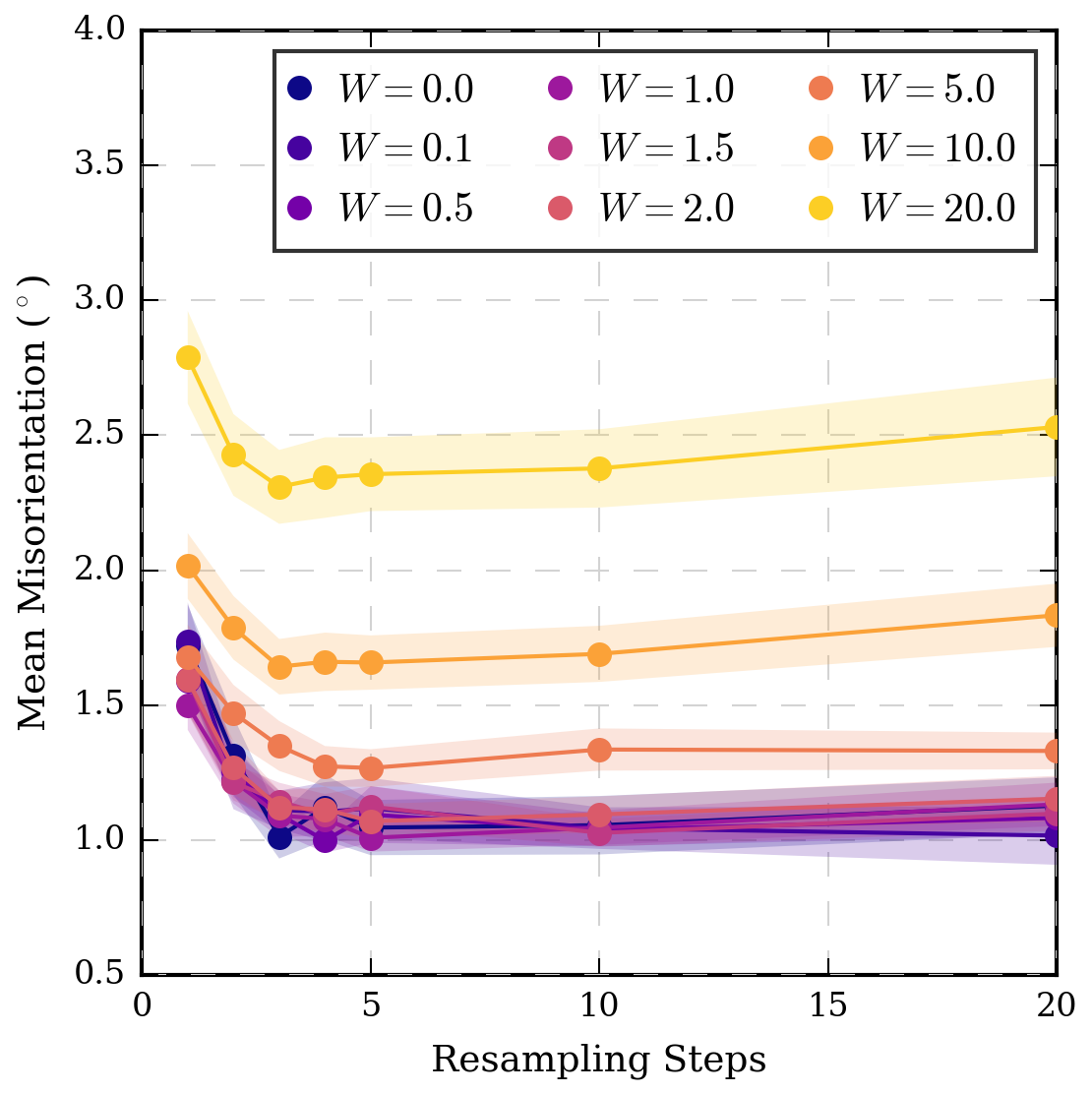}
  \caption{\textbf{Resampling Steps to Mitigate Internal Grain Misorientations.} The mean misorientation was estimated from the grain average orientation deviation (GROD), averaged over all grains in 1000 texture-conditioned synthesized microstructures across multiple resampling steps $N$. While an overemphasis of conditioning weight $w$ amplified grain misorientations, resampling steps eases these artifacts up to $N\approx5$.}
  \label{fig:nreps_vs_weight}
\end{figure}

\subsection{Internal Grain Misorientations}

Internal grain misorientations were present in all synthesized examples, a feature not present in the \textit{exaCA} dataset due to CA only sampling from discrete states. A continuous diffusion process is well-suited to generate natural images, which consist of smooth color transitions. Our synthetic dataset however, consists of constant plateaus of Euler angles within grains with sharp interfaces at grain boundaries. Internal grain misorientation are present in the generated images because a continuous process is used to approximate a discrete distribution. We are interested in exploring \textit{resampling} as a potential mitigation strategy. The reverse process follow continuous path discretized into $T$ time steps, hence slight numerical drifts can accumulate at each iteration degrading the fidelity of synthesized maps. Sequential resampling for noisy states allows for correction of these accumulated numerical drifts \cite{lugmayr2022repaint}. To test this, we generated a representative set of 1000 samples conditioned purely on texture. Using the Matlab open-source library MTEX \cite{bachmann2010texture}, we computed the Grain Reference Orientation Deviation (GROD) over all individual grains and the results shown in Fig. \ref{fig:nreps_vs_weight} correspond to the average misorientation across all grains. Our results indicate, that the density of internal grain misorientations were influenced by $N$ steps. Interestingly, increasing guidance weight amplified the mean GROD. Looking back at Fig. \ref{fig:process_fid}, this effect may contribute to the degradation of the FID score when $w >2$.

While resampling steps reduced the mean GROD up to $N=5$, misorientations could not be reduced below $1^\circ$. Perhaps a diffusion process natively defined on discrete classes may be better suited to generate microstructures from the \textit{exaCA} dataset \cite{austin2021structured}. While this is a limitation in the current study due to the nature of synthetic micrographs, internal grain misorientations exist in real EBSD maps. In fact, the presence of small misorientations is evidence of the ability of DDPMs to capture subtle low-angle gradients, features commonly observed in experimental datasets. 

\begin{figure}[tbp]
  \centering
  \includegraphics[width=1.0\linewidth, trim=0.0in 0.0in 0.0in 0.0in, clip]{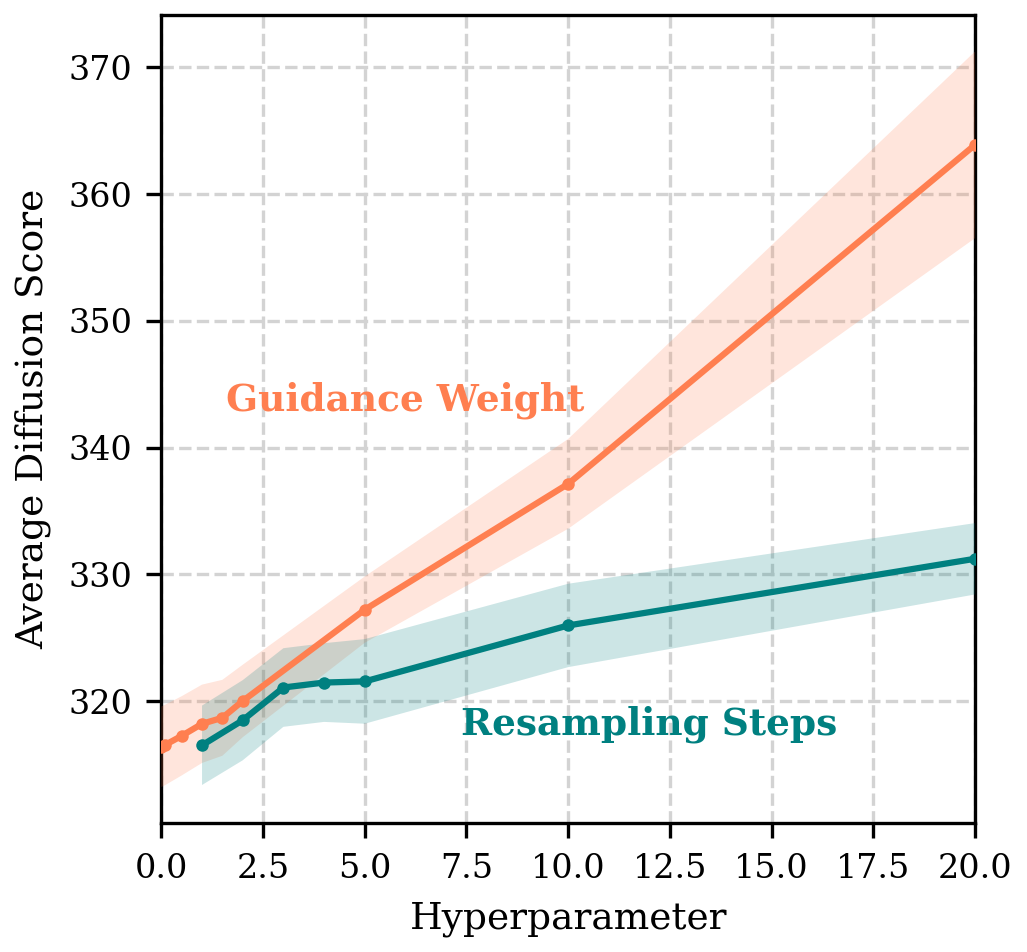}
  \caption{\textbf{Hyperparameters Effect on Diffusion Score.} Schematic illustrating the effect of the hyperparameters on the reverse process sampling. The diffusion score was computed by reversing diffusion from $t=300$ to $t=100$ over 1000 examples. Larger diffusion scores are indicative of out-of-distribution sampling}
  \label{fig:w_n_score}
\end{figure}

\subsection{Hyperparameter Tuning}

During inference, two tunable hyperparameters substantially influence the quality of the generated samples: guidance weight $w$, and resampling steps $N$. As previously mentioned, $w$ controls the emphasis on matching conditioning information, and $N$ controls number of resampling steps at each denoising step. In our process conditioning FID score vs guidance weight study we observed an optimal performance at $w\approx2.0$. Similarly, in Fig. \ref{fig:nreps_vs_weight}, we determined that resampling steps homogenize internal grain misorientations up until $N=5$ when generating texture-conditioned samples. In this section we examine the trade-offs associated with these hyperparameters.

Both hyperparameters have an effect on the diffusion score as observed in Fig. \ref{fig:w_n_score}, $w$ guides the sample towards regions where condition matching is probable, and $N$ provides additional samples to correct the numerical drift at the selected step. Although increasing both hyperparameters monotonically increase the average diffusion score, slight out-of-distribution sampling is not necessarily detrimental for generation. As indicated by the FID score analysis in Fig. \ref{fig:process_fid}, the optimal guidance weight was estimated to be $w \approx2$. Including conditioning information during the reverse diffusion process, narrows the distribution of plausible micrographs, hence, the larger score may reflect sampling from the sharper, narrower distribution rather than degradation in generation quality. 

The CFG expression for conditional generation described in Eq. {\ref{eqn:conditional_eqn}} can be rearranged as the following,

\begin{equation}
\begin{aligned}
\bm{\tilde{\epsilon}}_\theta\left(\mathbf{x}_t,t,\mathbf{c}\right) = \bm{\epsilon}_\theta\left(\mathbf{x}_t,t, \varnothing \right) + w\,(\bm{\epsilon}_\theta\left(\mathbf{x}_t,t,\mathbf{c}\right) - \bm{\epsilon}_\theta\left(\mathbf{x}_t,t, \varnothing \right))  
\label{eqn:cfg_eqn},
\end{aligned}
\end{equation}

\noindent it is evident that $w$ simply scales the distance between conditional and unconditional trajectories. Notably, $w$ shifts the trajectory without changing its magnitude, meaning that aggressive values can push the sampled step out-of-distribution significantly, which may degrade sample fidelity. Alternatively, small $w$ mitigates the signal from conditioning, and while sample fidelity is not necessarily degraded because the unconditional trajectory is always aiming toward high probable regions, conditioning matching is no longer guaranteed. The local minimum observed in our FID study showcases this inherent trade-off. Since the conditioning signal is sensitive to the provided contextual information, $w$ may have to be tuned for each individual case. 

The influence of $N$ is more subtle, within a single denoising step, the diffusion trajectory is updated multiple times by leveraging the context from sequential samples. Resampling is a mechanism to calibrate the diffusion steps by correcting accumulating numerical drifts. We found it mitigates certain artifacts such as internal grain misorientation. However, excessive resampling can over constrain the denoising process, and in some cases, push samples out-of-distribution as shown in Fig. \ref{fig:w_n_score}. Hence, carefully balancing $N$ is crucial to maximize sample fidelity while preserving diversity. 

\begin{figure}[tbp]
  \centering
  \includegraphics[width=1.0\linewidth, trim=0.0in 0.0in 0.0in 0.0in, clip]{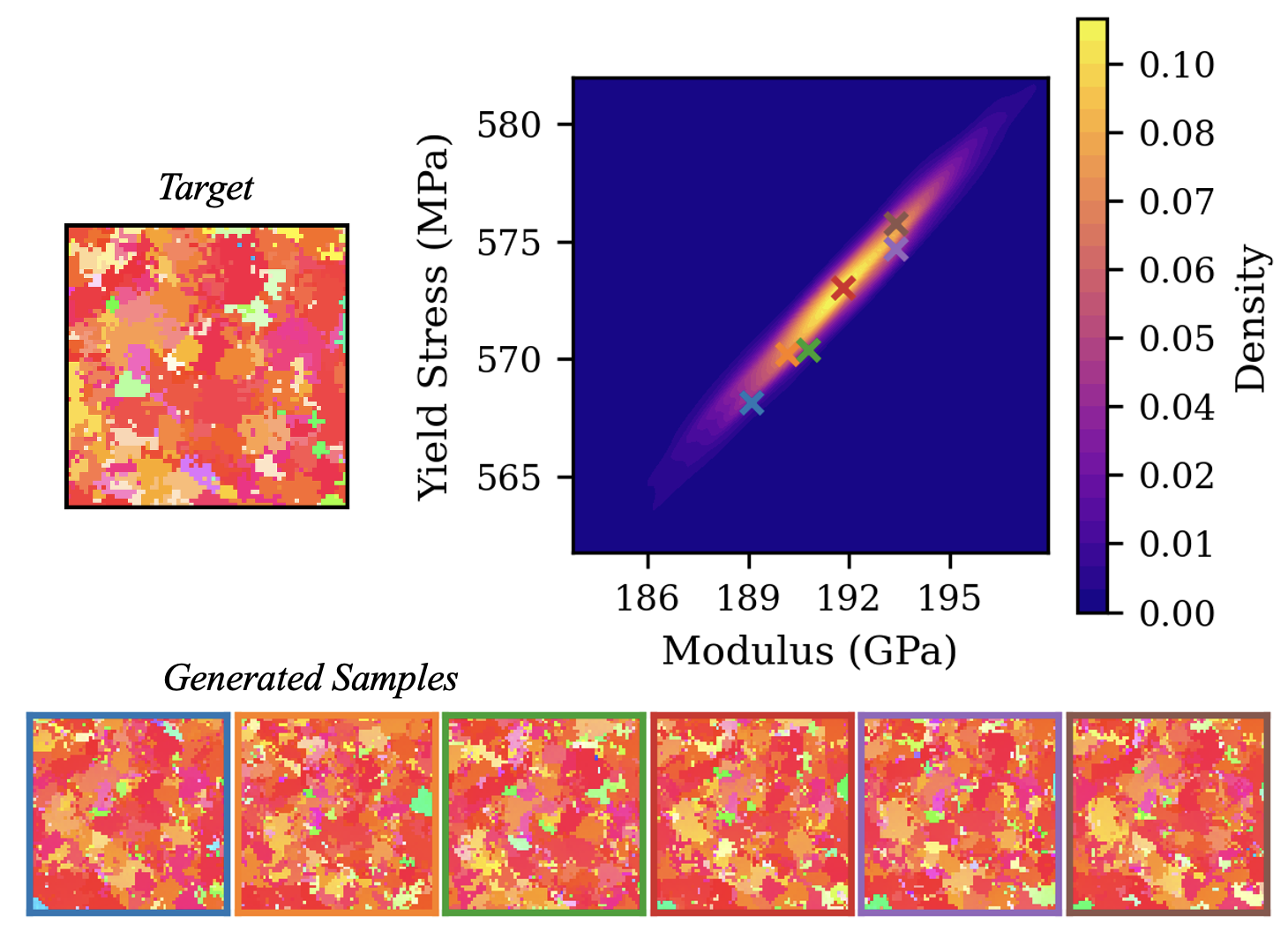}
  \caption{\textbf{DDPM Applied on Uncertainty Quantification.} A total of 1000 samples were generated conditionally with the texture information from the target microstructure with $w=3$ and $N=1$. The yield and elastic modulus for all generated samples were estimated with a surrogate model. The resulting distribution was estimated via Kernel Density Estimation (KDE).}
  \label{fig:uq_example}
\end{figure}

\subsection{Use Case: Uncertainty Quantification}

In the following section we demonstrate a use-case scenario of our DDPM as a tool for uncertainty quantification (UQ). Traditional deterministic models are insufficient to capture localized behaviors that drive failure because the microstructure of a polycrystalline material is inherently stochastic. Therefore, a statistical ICME framework that models microstructural variability is required. By incorporating this DDPM into an ICME framework, researchers can evaluate the mechanical properties of statistically equivalent microstructures and leverage the simulated results to explicitly determine confidence intervals rather than a single deterministic value as visualized in Figure \ref{fig:uq_example}. To demonstrate this, we conditionally generated 1000 samples that follow the same texture as a target microstructure. For a rapid demonstration of this framework, rather than running CPFEM simulations on all samples, we opted to use a pre-trained surrogate model to predict the yield stress and elastic modulus of each sample. The surrogate model, trained on the same dataset as the DDPM, consists of a series of convolutional layers that take in $64 \times64 \times 3$ tensor of Euler angles, followed by an MLP that outputs the predicted yield stress and elastic modulus. This is the exact same model previously used to evaluate property conditioning on Fig. \ref{fig:conditional_property}. For the texture corresponding to the target micrograph, based on variability in structure produced by the DDPM and assuming that the error from the surrogate is negligible compared to the structural variation, the predicted property variation shown in the contour map in Fig. \ref{fig:uq_example} is approximately 191.7 +/- 2.9 GPa for the uniaxial elastic modulus and  572.8 +/- 4.5 MPa for the yield strength. Note that in this particular example the variation is rather small, this is because the elastic modulus and yield are aggregate volume-averaged quantities. The entire microstructural area contributes to this single number and spatial arrangement of the structure is relatively unimportant. However, for applications such as fatigue, where failure is dominated by stochasticity in the local structure, for example, a particularly susceptible grain boundary between a soft orientated grain next to a hard oriented grain, this form of UQ analysis is even more critical.

\section{Conclusion}

In this work we establish DDPMs as expressive generative models for orientation-imaging-based microstructure synthesis and inverse design tasks. We demonstrate this capability through a case study focused on AM microstructures. The generated samples capture complex key features characteristic of AM materials and improve upon traditional approaches based on sampling from statistical representations. We explore a classifier-free conditional generation strategy across multiple features prescribing process settings, desired properties, and quantitative statistical descriptors. We show that tuning of several inference hyperparameters could improve generative quality to a degree but some artifacts (internal misorientations not present in the training dataset) remained. This work highlights the potential of AI-based generative modeling for materials inverse design and microstructure sampling workflows relevant to AM materials qualification and certification.

\section*{Acknowledgments}
PFZ and AZ acknowledge support from the AI Initiative under the Laboratory Directed Research and Development (LDRD) Program at Oak Ridge National Laboratory (ORNL). PLF, JM, and PFZ acknowledge support from the U.S. Department of Energy’s Advanced Materials and Manufacturing Technologies Office (AMMTO), Office of Critical Minerals and Energy Innovation (CMEI). ORNL is managed by UT-Battelle, LLC, for the U.S. Department of Energy under Contract No. DE-AC05-00OR22725. PLF acknowledge support by appointment to the Education Collaboration at Oak Ridge National Laboratory Program (ECO), sponsored by the U.S. Department of Energy and administered by the Oak Ridge Institute for Science and Education.

\FloatBarrier
\label{references}
\bibliographystyle{unsrtnat}
\bibliography{refs}

\clearpage 
\appendix

\setcounter{page}{1}
\setcounter{section}{0}
\setcounter{figure}{0}
\setcounter{table}{0}

\renewcommand{\thepage}{S\arabic{page}}
\renewcommand{\thesection}{S\arabic{section}}
\renewcommand{\thefigure}{S\arabic{figure}}
\renewcommand{\thetable}{S\arabic{table}}

\section{Supplemental Materials}

\begin{figure}[htbp]
  \centering

  \includegraphics[width=1.0\linewidth]{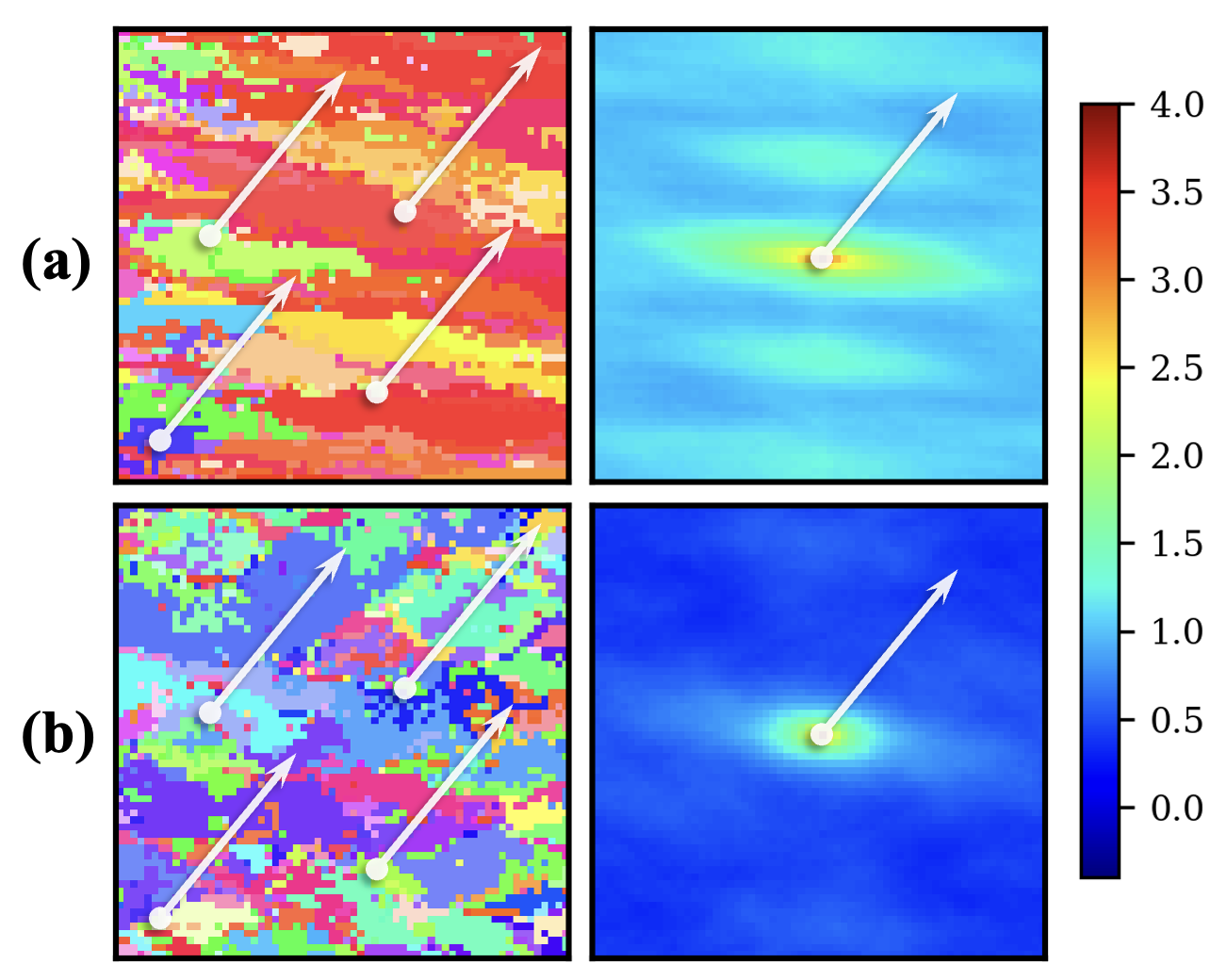}
  \caption{\textbf{Autocorrelation Map Interpretation.} The white arrow corresponds to $t$ from Eq. \ref{eqn:two-point}. The correlation between material state $\mathbf{g}$ at the tail $(\mathbf{x})$ and the state $\mathbf{g}^\prime$ at the head $(\mathbf{x}+t)$ of a fixed $t$ is evaluated by sweeping over all possible positions $\mathbf{x}$ within the EBSD micrograph. The spatial autocorrelation measures the likelihood of staying in the same state when moving away a distance $t$. The center of the autocorrelation map corresponds to $t=0$, which is always the maximum value because the head and the tail land on identical material states ($\mathbf{g} = \mathbf{g}^\prime$). Any distance and direction away from this center point corresponds to the autocorrelation value at the respective length and orientation of $t$. Morphology is described by short-range correlations at the center of the map, capturing the average shape, size, and aspect ratio of features across the micrograph. Example (a) is a textured material with horizontal columnar grains, short-range correlations capture the overall shape as an ellipse. Example (b) is an untextured materal with equiaxed-like grains, in this case short-range correlations form a circular shape..}
  \label{fig:ss_example}
\end{figure}

\begin{figure}[htbp]
  \centering

  \includegraphics[width=1.0\linewidth]{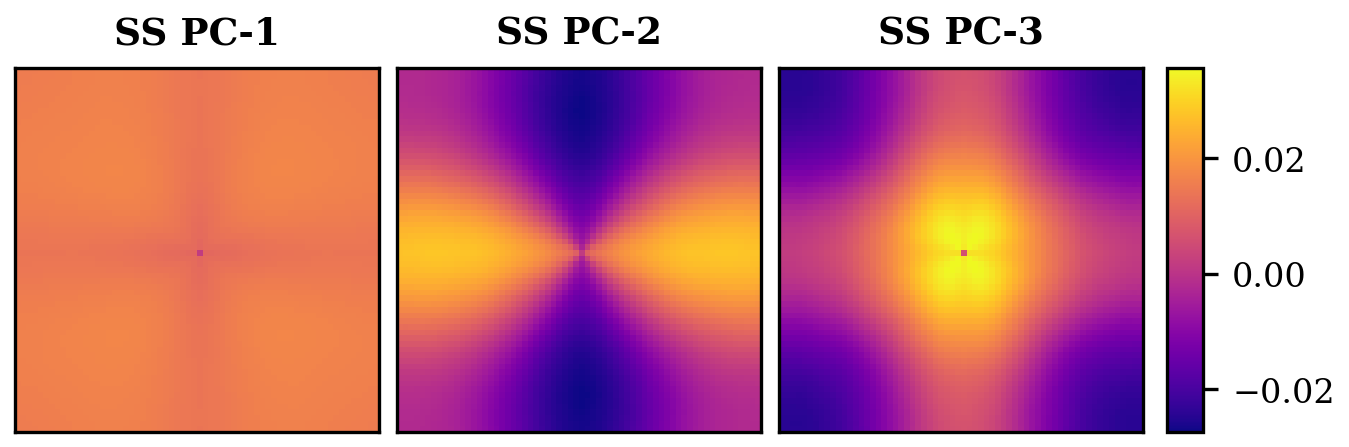}
  \caption{\textbf{Spatial Statistics Principal Axes.} PCA identifies principal axes along which the spread of the data is maximized when projected onto them. A Principal Component (PC) score is then computed as the sum of all data features weighted by these principal axes. SS PC-1, SS PC-2 and SS PC-3 correspond to the first three principal axes. Plotting the principal axes reveal the unique morphological feature captured by each component. For PC-2 vertical and horizontal bands are heavily weighted, hence encoding grain aspect ratio information. For PC-3 the region closest to the center is heavily weighted, encoding local grain size information. PC-1 weights do not reveal a shape pattern, suggesting that a different feature not necessarily representative of morphology is being captured.}
  \label{fig:ss_pca_v}
\end{figure}

\begin{figure}[htbp]
    \centering
    \begin{subfigure}[b]{0.48\textwidth}
        \centering
        \includegraphics[width=\textwidth]{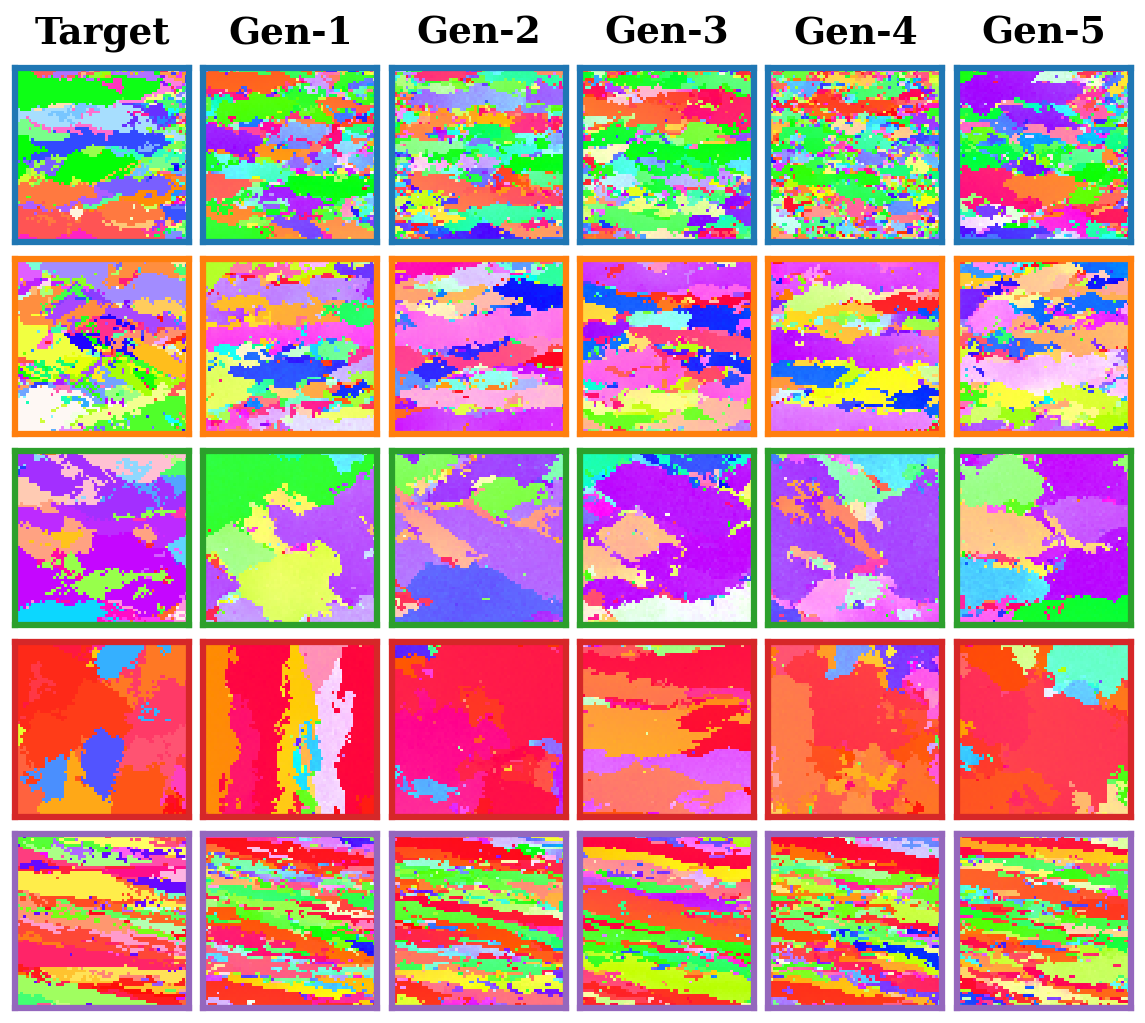}
        \caption{}
        \label{fig:ipf_ss_pc1_a}
    \end{subfigure}

    \vspace{0.5cm} 
    \begin{subfigure}[b]{0.48\textwidth}
        \flushleft
        \includegraphics[width=\textwidth]{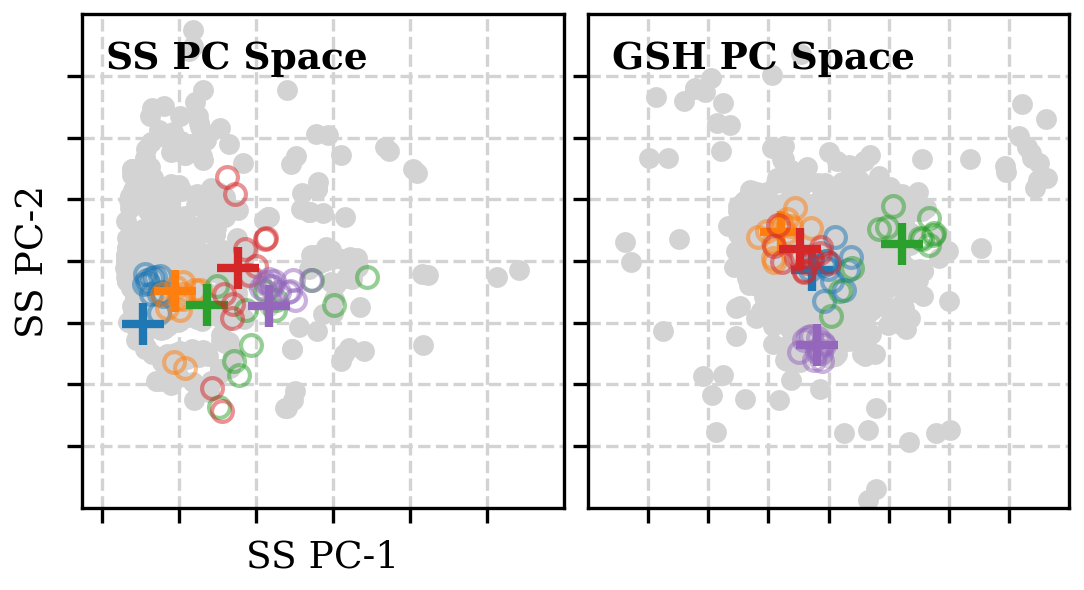}
        \caption{}
        \label{fig:ipf_ss_pc1_b}
    \end{subfigure}

    \caption{\textbf{SS PC-1 Diagnostics with Texture-conditioned Samples}.  (a) IPF-Z maps of generated and target samples conditioned only on texture. (b) SS and GSH PC diagnostics plots. Despite no morphology information provided, the generated samples agree with the target PC representation.}
    \label{fig:ss_pc1}
\end{figure}

\subsection{Dimensionality Reduction of Spatial Statistics.}

A compressed representation of the $64 \times 64$ autocorrelation maps, via PCA, was used to diagnose generated samples morphological agreement with target examples. For an explanation of autocorrelation maps refer to Fig. \ref{fig:ss_example}. The first three principal axes capture morphological features as illustrated in Fig. \ref{fig:ss_pca_v}. As described in Fig. \ref{fig:ss_example}, short-range correlations, located around the center of the map, capture the average grain morphology. PC-3 emphasizes variance at the center, thus capturing grain size. PC-2 emphasizes vertical and horizontal bands emerging from the center, meaning that grain elongation is being captured. Since our augmentation strategy consist of $90^\circ$, $180^\circ$, and $270^\circ$ rotations, columnar-grained structures are either horizontal or vertical, which explains the cross-like shape of PC-2. The morphological features captured by PC-1 are not immediately obvious. Its uniform weighing across the map suggest an emphasis that extends beyond short-range correlations. Long-range correlations, where $t$ is large, indicate a similarity of crystallographic orientations that can span across multiple grain boundaries, which is a common characteristic in textured materials. Observe Fig. \ref{fig:ss_example} , the long range autocorrelation values are higher for the highly textured example (a) compared to the untextured example (b). Therefore PC-1, which emphasizes long-range correlations, captures texture information. To confirm our intuition we sampled five examples with unique SS PC-1 and conditionally generated samples with only texture information. Generated samples as well as PC space diagnostic plots are shown in Fig. \ref{fig:ss_pc1}. As anticipated, the SS PC-1 of the generated samples successfully matched the target. 

\begin{figure}[tbp]
  \centering
  \includegraphics[width=1.0\linewidth, trim=0.0in 0.0in 0.0in 0.0in, clip]{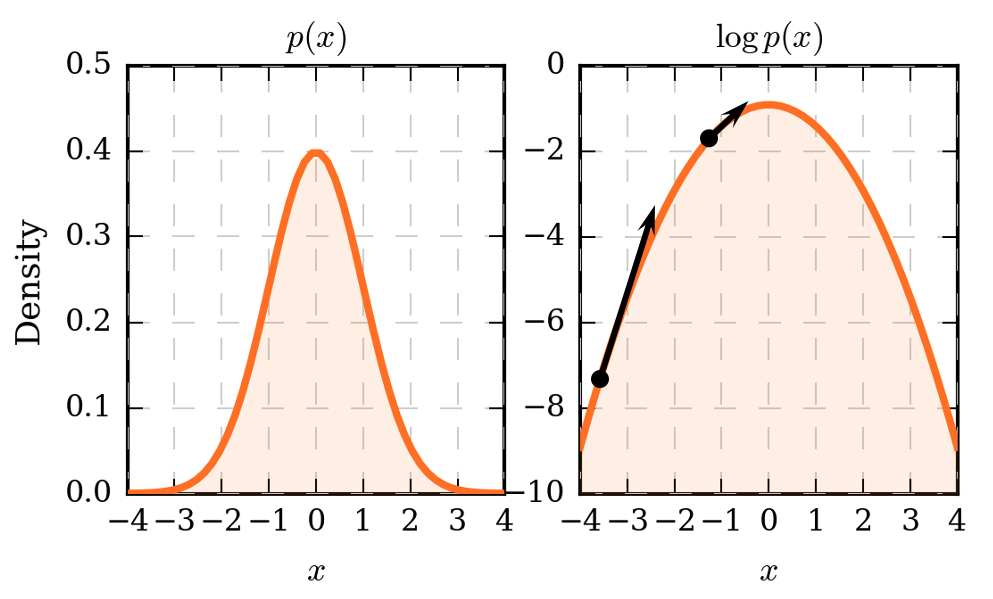}
  \caption{\textbf{Score Function Visualization.} Schematic illustrating the score function $s(x) = \nabla \log{p(x)}$ on a standard normal distribution. The magnitude of the arrows in the right figure represent the score, hence the score is minimized as $x$ approaches local mode of the distribution}
  \label{fig:score_example}
\end{figure}
\subsection{Score Function}

Score-based diffusion models learn trajectories in data space that guide random noise toward highly probable samples. The direction in which a sample must move is calculated directly by the \textit{score function}. In contrast, DDPMs predicts the injected noise at each time step rather than estimating the score vector field directly. As seen in equation \ref{eqn:diffusion_score}, this predicted noise is proportional to the score, hence, the score function intuition is relevant to understand the underlying mechanism behind DDPMs. Mathematically, the score function is $\nabla_{x_t} \log p(x_t)$, where $p(x_t)$ is the data distribution at a noisy state $t$. An intuitive visualization of the score function on a normal distribution is shown in \ref{fig:score_example}. The further a sampled noisy state $x_t$ deviates is from a local mode of the distribution $p(x_t)$, the greater the magnitude of the score vector, resulting in a strong pull towards high density regions.

\subsection{Additional Cross-conditioning Experiments.}

\begin{figure}[tbp]
  \centering
 
  \includegraphics[width=1.0\linewidth, trim=0.0in 0.0in 0.0in 0.0in, clip]{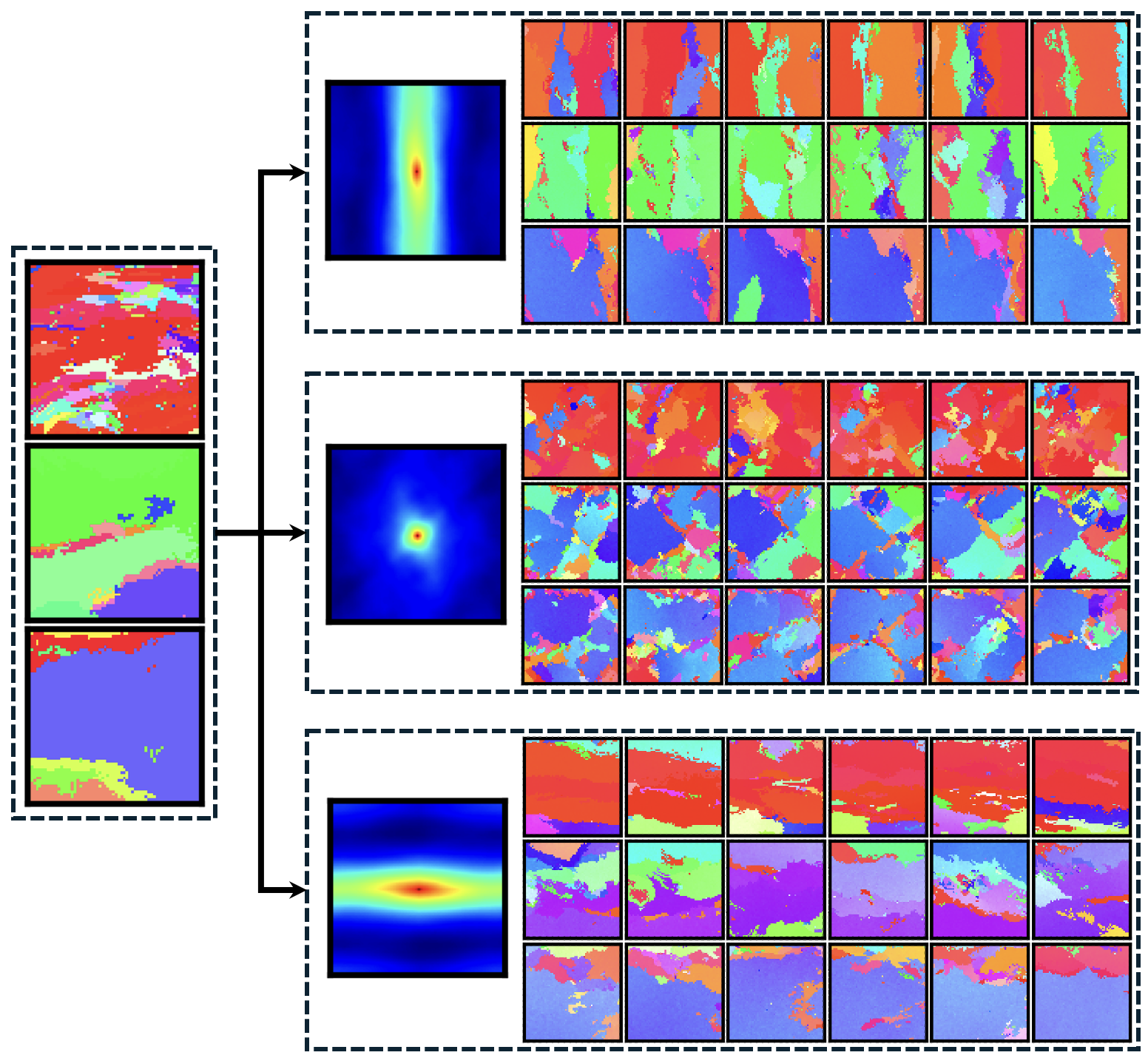}
  \caption{\textbf{Texture and Morphology Generation.} Generating out-of-distribution data by combining three samples from the validation dataset (left-most examples) with three unique morphologies. Samples were generated with $w=3.0$ and $N=1$}
  \label{fig:ood_texture_morph}
\end{figure}

As further evaluation of the DDPM's capacity to synthesize novel microstructure, we generated samples from clearly out-of-distribution combinations of textures and morphologies. In Fig. \ref{fig:ood_texture_morph}, three distinct textures that approximately resemble an $<001>$ , $<011>$ , and $<111>$ fiber were extracted from the validation dataset. Similarly, three representative morphologies corresponding to horizontal-columnar, vertical-columnar and equiaxed-grained micrographs were selected. Notably, although $<011>$ fibers were uncommon in our dataset, the DDPM successfully generated a microstructure by combining both vertical columnar grains and $<011>$ fiber texture. For the other two morphology conditions, however, the $<011>$  fiber was only weakly represented. The $<111>$ fiber example, particularly under columnar morphology conditions, the generated microstructure resulted in one large grain with the target texture rather than fully reproducing the overall morphology. 

\begin{figure}[tbp]
  \centering

  \includegraphics[width=1.0\linewidth, trim=0.0in 0.0in 0.0in 0.0in, clip]{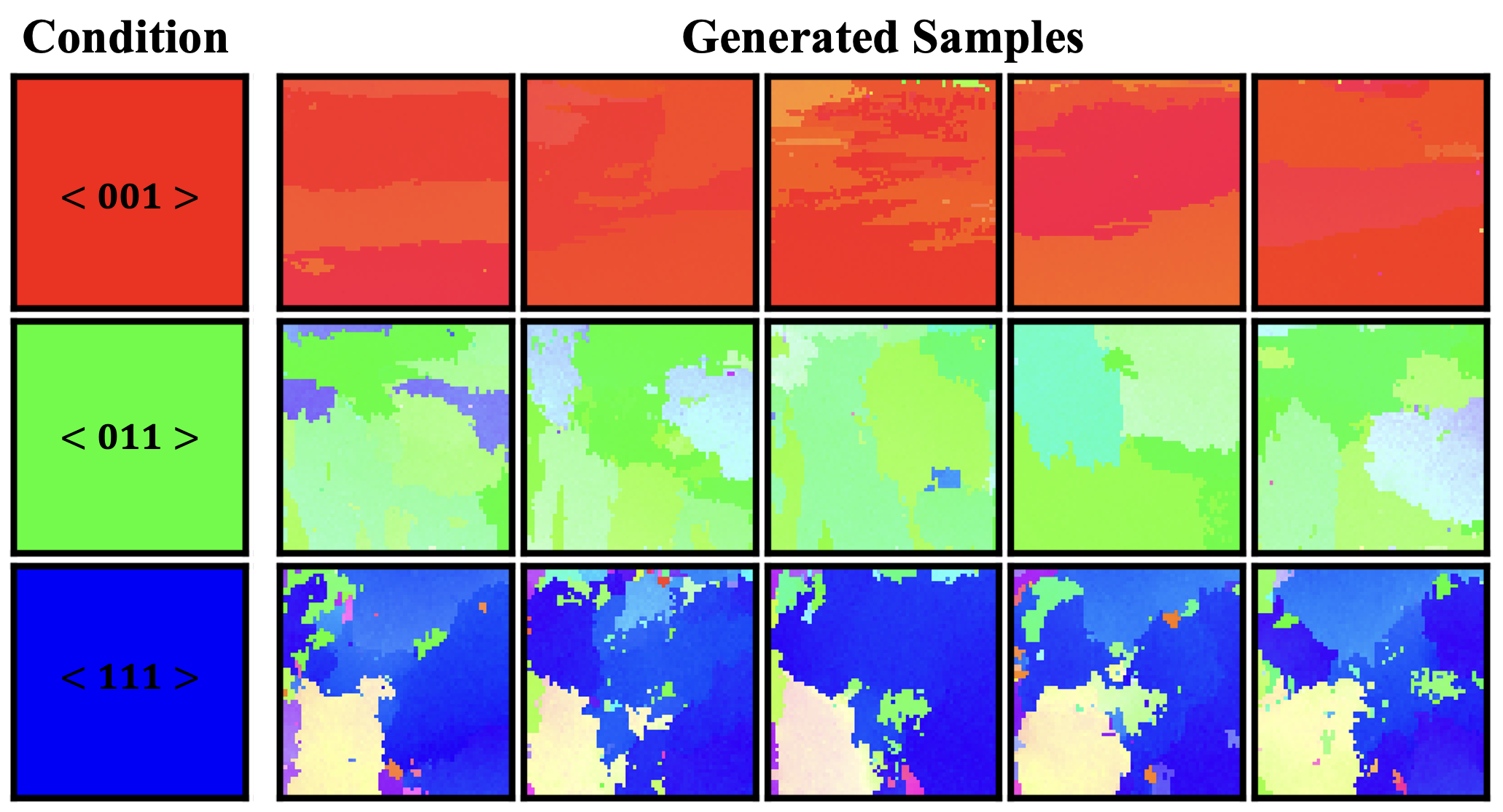}
  \caption{\textbf{SX-conditioned Generation.} Generating samples from $<001>$, $<011>$, and $<111>$ single-crystal (SX) texture data. Samples were generated with $w=3.0$ and $N=1$}
  \label{fig:ood_texture}
\end{figure}

The second experiment consisted on texture conditioned generation using $<001>$ , $<011>$ , and $<111>$ single-crystal textures, which were clearly not previously seen by the DDPM.  Remarkably, the DDPM was still able to generate structures that partially resembled the target textures while at the same time preserving some morphological features observed in the validation dataset. 

\begin{figure}[tbp]
  \centering

  \includegraphics[width=1.0\linewidth, trim=0.0in 0.0in 0.0in 0.0in, clip]{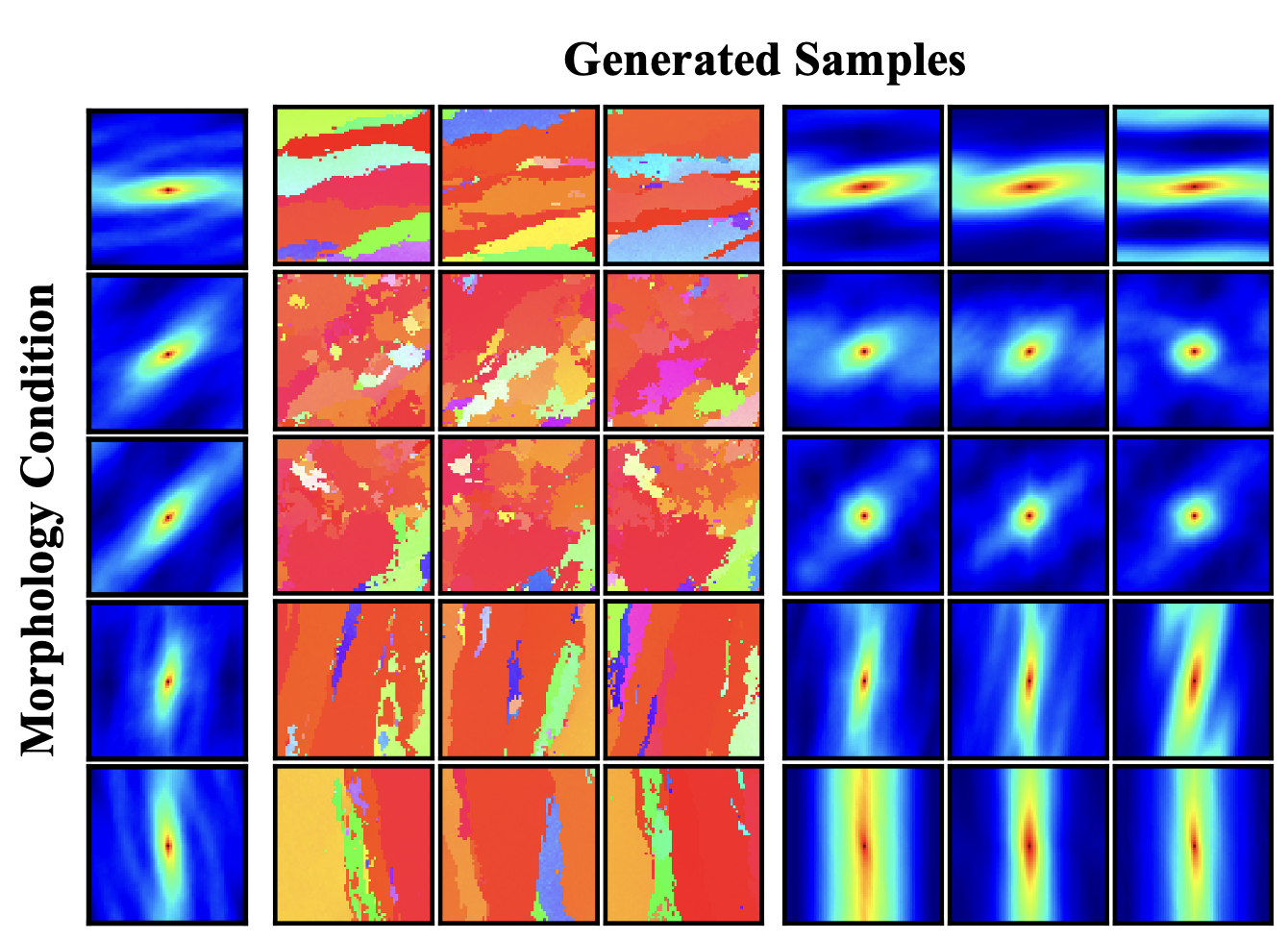}
  \caption{\textbf{Rotated Morphology Generation.} Generating samples from rotated columnar autocorrelation maps, rotated by $0^\circ$, $22.5^\circ$, $45^\circ$, $67.5^\circ$, and $90^\circ$ from top to bottom row. Samples were generated with $w=3.0$ and $N=1$}
  \label{fig:ood_morph}
\end{figure}

The third experiment, shown in Fig. \ref{fig:ood_morph}, consisted of physically rotating a columnar-grained micrograph obtained from the validation dataset and conditioning the DDPM to generate rotated variants of this morphology. The most notable case is the $45^\circ$ rotation, for which the DDPM appeared to compensate for the diagonal long-range correlations in the autocorrelation map by increasing the grain size which partially approximates the target morphology. 

\end{document}